\documentclass[authoryear,10pt,letterpaper]{elsarticle}
\usepackage[left=3cm,right=3cm,top=3cm,bottom=3cm]{geometry}

\usepackage{amssymb,amsmath,verbatim,mathtools,needspace,enumitem,etoolbox,graphicx,physics,microtype,afterpage}
\usepackage{gensymb, soul}
\usepackage[dvipsnames, usenames]{xcolor}

\definecolor{linkcolor}{rgb}{0.0,0.3,0.5}
\usepackage[unicode,colorlinks=true, linkcolor=linkcolor, citecolor=linkcolor, filecolor=linkcolor,urlcolor=linkcolor]{hyperref}

\usepackage{titlesec}

\setcitestyle{aysep={}}

\usepackage{setspace}
\usepackage{tocloft}

\journal{iScience}
\usepackage{tensor}     % manipulate tensors
\usepackage{bm}         % \bm{<text>} Bold math symbols
\usepackage[utf8]{inputenc} % accented characters for .bib file using XeLaTex
\usepackage[section]{placeins} % figures processing

\newcommand{\nc}{\newcommand*} 

\nc{\al}{\alpha}
\nc{\s}{\sigma}
\nc{\dt}{\delta}
\nc{\Dt}{\Delta}
\nc{\Ld}{\Lambda}
\nc{\p}{\partial}
\nc{\om}{\omega}
\nc{\Om}{\Omega}
\nc{\rd}{\mathrm{d}}
\nc{\Od}[1]{\mathcal{O}(#1)} % order operator
\nc{\kp}{\kappa}
\nc{\one}{\uppercase\expandafter{\romannumeral1}}
\nc{\two}{\uppercase\expandafter{\romannumeral2}}
\nc{\three}{\uppercase\expandafter{\romannumeral3}}
\def\({\left(}
\def\){\right)}
\def\[{\left[}
\def\]{\right]}
\def\e{\begin{equation}}
\def\q{\end{equation}}
\def\m{\begin{eqnarray}}
\def\n{\end{eqnarray}}
\nc{\Eq}[1]{Eq.~\eqref{#1}}     % equation
\nc{\Fig}[1]{Fig.~\ref{#1}}     % figure
\nc{\Table}[1]{Table~\ref{#1}}  % table
\nc{\Sec}[1]{Sec.~\ref{#1}}     % section
\nc{\Msun}{M_\odot}             % solar mass
\nc{\fpbh}{f_{\mathrm{pbh}}}    % f_pbh
\nc{\fpbhn}{f_{\mathrm{pbh0}}}    % f_pbh
\nc{\mR}{\mathcal{R}} % merger rate density
\nc{\seq}{\sigma_{\mathrm{eq}}}
\nc{\ogw}{\Omega_{\mathrm{GW}}}
\nc{\gpcyr}{\mathrm{Gpc}^{-3}\,\mathrm{yr}^{-1}}
\nc{\lvc}{LIGO/Virgo} % LIGO-VIRGO collaboration
\nc{\SNR}{\mathrm{SNR}} % signal to noise ratio
\nc{\mmin}{{m_{\mathrm{min}}}}
\nc{\mmax}{{m_{\mathrm{max}}}}
\nc{\Mmin}{{M_{\mathrm{min}}}}
\nc{\fmin}{{f_{\mathrm{min}}}}
\nc{\VT}{\mathrm{VT}}
\nc{\rhoGW}{\rho_{\mathrm{GW}}}
\nc{\vth}{\vec{\theta}}
\nc{\vd}{\vec{d}}
\nc{\vla}{\vec{\lambda}}
\nc{\Nobs}{N_{\mathrm{obs}}}
\nc{\av}[1]{\langle #1 \rangle} % average bracket
\nc{\km}{\mathrm{km}}
\nc{\Mpc}{\mathrm{Mpc}}
\nc{\Tobs}{T_{\mathrm{obs}}}
\nc{\Ntemp}{N_{\mathrm{temp}}}
\nc{\addref}{[\textcolor{red}{add ref}] } % placeholder of references
\nc{\eg}{\textit{e.g.~}}
\nc{\app}{\approx}
\nc{\hf}{\frac{1}{2}}
\nc{\discuss}{\textcolor{red}{Add discussion here!}}
\nc{\red}[1]{\textcolor{red}{#1}}
\nc{\mH}{\mathcal{H}}
\nc{\cs}{c_s^2}
\nc{\Sij}[1]{S_{ij}^{(#1)}}
\nc{\vi}[1]{v_i^{(#1)}}
\nc{\no}{\nonumber}
\def\<{\left\langle}
\def\>{\right\rangle}

\def\half{{1\over 2}}
\nc{\bk}{\bm{k}}
\nc{\bq}{\bm{q}}
\nc{\bp}{\bm{p}}
\nc{\bl}{\bm{l}}
\nc{\bx}{\bm{x}}
\nc{\be}{\mathbf{e}}
\nc{\mS}{\mathcal{S}}
\nc{\te}{\tilde{\eta}}
\nc{\tp}{\tilde{p}}
\nc{\tk}{\tilde{k}}
\nc{\tx}{\tilde{x}}
\nc{\tF}{\tilde{F}}
\nc{\tA}{\tilde{A}}
\nc{\mkpq}{|\bk-\bp-\bq|}
\nc{\mpq}{|\bp-\bq|}
\nc{\mkp}{|\bk-\bp|}
\nc{\mSi}[1]{\mS^{(#1)}({\bk, \eta})}
\nc{\vk}{\vec{k}}
\nc{\kstar}{k_*}
\nc{\xstar}{x_*}
\nc{\mpbh}{m_{\rm{pbh}}}
\nc{\Ci}{\mathrm{Ci}}
\nc{\Si}{\mathrm{Si}}
\nc{\fnl}{f_\mathrm{NL}}
\nc{\gnl}{g_\mathrm{NL}}
\nc{\Fnl}{F_\mathrm{NL}}
\nc{\Gnl}{G_\mathrm{NL}}
\nc{\togw}{\tilde{\Omega}}
\nc{\md}{\mathrm{d}^3}
\renewcommand{\vec}[1]{\boldsymbol{#1}} % Uncomment for BOLD vectors.

\begin{document}
%%%%%%%%%%%%%%%%%%%%%%%%%%%%%%%%%%%%%%%%%%%%%%%%%%%%%%%%%%%%%%%%%%%%%%%%%%%%%%%%
{\LARGE A topic review on probing primordial black hole dark matter with scalar induced gravitational waves}

\bigskip
%%%%%%%%%%%%%%%%%%%%%%%%%%%%%%%%%%%% author %%%%%%%%%%%%%%%%%%%%%%%%%%%%%%%%%%%%

\renewcommand{\thefootnote}{$\dagger$}
{Chen Yuan$^{1,2,}$}\footnote{yuanchen@itp.ac.cn} 
\renewcommand{\thefootnote}{$\ddagger$}
{Qing-Guo Huang$^{1,2,3,}$}\footnote{huangqg@itp.ac.cn}

\medskip

{\small   \it
	{$^1$School of Physical Sciences, 
		University of Chinese Academy of Sciences, 
		No. 19A Yuquan Road, Beijing 100049, China}}

\medskip

{\small \it {$^2$CAS Key Laboratory of Theoretical Physics, 
		Institute of Theoretical Physics, Chinese Academy of Sciences,
		Beijing 100190, China}	
}

\medskip

{\small \it {$^3$Center for Gravitation and Cosmology, 
		College of Physical Science and Technology, 
		Yangzhou University, Yangzhou 225009, China}}

%%%%%%%%%%%%%%%%%%%%%%%%%%%%%%%%%%%%% date %%%%%%%%%%%%%%%%%%%%%%%%%%%%%%%%%%%%%
%\date{\today}
	
%%%%%%%%%%%%%%%%%%%%%%%%%%%%%%%%% abstract %%%%%%%%%%%%%%%%%%%%%%%%%%%%%%%%%%%%%
\section*{abstract}

Primordial black holes (PBHs) might form from the collapse of over-densed regions generated by large scalar curvature perturbations in the radiation dominated era. Despite decades of various independent observations, the nature of dark matter (DM) remains highly puzzling. Recently, PBH DM have aroused interest since they provide an attracting explanation to the merger events of binary black holes discovered by LIGO/VIRGO and may play an important role on DM. During the formation of PBH, gravitational waves will be sourced by linear scalar perturbations at second-order, known as the scalar-induced gravitational waves (SIGWs), which provides a new way to hunt for PBH DM. This topic review mainly focus on the physics about SIGWs accompanying the formation of PBH DM.
	
%\pacs{???}
	
%\maketitle
	
%%%%%%%%%%%%%%%%%%%%%%%%%%%%%%%%%%%%%%%%%%%%%%%%%%%%%%%%%%%%%%%%%%%%%%%%%%%%%%%%

\tableofcontents

\newpage

\section{Introduction} Primordial black holes (PBHs) have aroused interest recently, not only because they can represent the dark matter (DM) in our Universe, but also can explain the binary black hole mergers events \cite{Sasaki:2016jop,Chen:2018czv,Raidal:2018bbj,DeLuca:2020qqa,Hall:2020daa,Bhagwat:2020bzh,Hutsi:2020sol} discovered by LIGO \cite{Abbott:2016blz} if $\sim 10^{-3}$ DM is in the form of PBHs. So far, there is no evidence for PBHs. Although various observations have constrained the fraction of DM in the form of PBHs \cite{Tisserand:2006zx,Carr:2009jm,Barnacka:2012bm,Griest:2013esa,Graham:2015apa,Brandt:2016aco,Chen:2016pud,Wang:2016ana,Gaggero:2016dpq,Ali-Haimoud:2016mbv,Blum:2016cjs,Horowitz:2016lib,Niikura:2017zjd,Zumalacarregui:2017qqd,Nakama:2017qac,Abbott:2018oah,Magee:2018opb,Chen:2018rzo,Niikura:2019kqi,Chen:2019irf,Authors:2019qbw,Wang:2019kzb}, $\fpbh$, there still exist an open window in the mass range of $[10^{-16},10^{-14}] \cup [10^{-13},10^{-12}] M_\odot$ where PBHs are possible to present all the DM in our Universe. 

PBH is an old conception and it can date back to 1974 when Hawking and Carr proposed that black holes can be generated due to the collapse of over-densed regions in the early universe \cite{Carr:1974nx,Carr:1975qj}. 
The formation of PBHs is a threshold process.
Once scalar perturbations exceed a critical value, they would generate an over-densed region which would immediately undergo gravitational collapse to form a single PBH when the comoving size of such region is of the order of the horizon size.
The exact calculation of the PBH mass function, $\beta$, which describes the mass fraction of the Universe contained within PBHs at the formation time is still a debating and complicated question by today. 

Among all the constraints on PBH DM, the scalar-induced gravitational waves (SIGWs) provide a quite stringent constraint which can be several orders of magnitude better than the other constraints \cite{Chen:2019xse} in a certain mass range of PBHs. During radiation dominant (RD) epoch, scalar perturbations will alter the quadrupolar moment of the radiation and thus emit GWs at second-order \cite{tomita1967non,Matarrese:1992rp,Matarrese:1993zf,Matarrese:1997ay,Noh:2004bc,Carbone:2004iv,Nakamura:2004rm}.
Therefore, SIGWs were inevitably generated during the formation of PBHs, providing a powerful tool to hunt for PBH DM.
Moreover, PBHs are generated by large perturbations at small scales much larger than those on CMB scales, the second-order GWs induced by the enhanced perturbations sourced by the linear perturbations may exceed the first-order tensor inflationary modes \cite{Saito:2008jc}. See more relevant studies for SIGWs in \cite{Ananda:2006af,Baumann:2007zm,Saito:2008jc,Arroja:2009sh,Assadullahi:2009jc,Bugaev:2009kq,Bugaev:2009zh,Saito:2009jt,Bugaev:2010bb,Alabidi:2013lya,Nakama:2016enz,Nakama:2016gzw,Inomata:2016rbd,Orlofsky:2016vbd,Garcia-Bellido:2017aan,Sasaki:2018dmp,Espinosa:2018eve,Kohri:2018awv,Cai:2018dig,Bartolo:2018evs,Bartolo:2018rku,Unal:2018yaa,Byrnes:2018txb,Inomata:2018epa,Clesse:2018ogk,Cai:2019amo,Inomata:2019zqy,Inomata:2019ivs,Cai:2019elf,Yuan:2019udt,Cai:2019cdl,Lu:2019sti,Yuan:2019wwo,Tomikawa:2019tvi,DeLuca:2019ufz,Yuan:2019fwv,Inomata:2020tkl,Inomata:2020yqv,Inomata:2020lmk,Yuan:2020iwf,Papanikolaou:2020qtd,Zhang:2020ptw,Kapadia:2020pnr,Zhang:2020uek,Domenech:2020ssp,Dalianis:2020gup,Atal:2021jyo}. 
The SIGWs from PBHs are first calculated by Saito and Yokoyama in \cite{Saito:2008jc} where they evaluate the energy density of SIGWs from monochromatic PBHs. They found that SIGWs from the current PBH DM in our Universe could be detected by pulsar timing arrays and space-based GW detectors. After the detection of GWs, intriguing studies emerged in this field in the recent years and there are hundreds of studies concerning SIGWs from PBHs so far.

Since there are a lot of reviews on PBHs in literature, e.g. some recent reviews given in \cite{Carr:2020gox,Green:2020jor,Carr:2020xqk}, we mainly focus on SIGWs inevitably generated during the formation of PBHs in this paper. This paper will be organized as follows. In Sec.~II, we give a brief introduction to the formation of PBHs. The physics about the SIGWs will be reviewed in Sec.~III and then we discuss how to use the SIGWs to probe PBHs in Sec.~IV. Finally, summary and outlook are given in Sec.~V. 

%In this review, we will discuss the SIGWs inevitably generated during the formation of PBHs. We will introduce the basic calculation of SIGWs and relevant works in this field. Since we mainly focus on SIGWs, we will only take a brief introduction to the PBH formation mechanism. The latest constraints on PBH DM has been discussed in two recent review \cite{Carr:2020gox,Green:2020jor}. For the recent development in PBHs, we recommend the readers to a recent review \cite{Carr:2020xqk}.

\section{Formation of Primordial Black Holes}
In this section, we will introduce the formation of PBHs and take a brief review on calculating the mass function of PBHs, $\beta$.
PBHs are generated from the collapse of all the matter inside the Hubble volume. Therefore, there exists a one-to-one correspondence between the mass of PBHs and the comoving frequency $f_*$, namely \cite{Carr:1974nx,Carr:1975qj}
\e\label{mpbh}
{m_{\mathrm{pbh}}^*}\approx2.3\times10^{18}\Msun\left(\frac{3.91}{g_*^{\mathrm{form}}}\right)^{1 / 6}\left(\frac{H_{0}}{f_*}\right)^{2},
\q
where $g_*^{\mathrm{form}}$ is the corresponding degrees of freedom and $H_0$ is the Hubble constant by today.
Moreover, the one-to-one correspondence can be transferred to another useful form, namely
\e\label{mpbh_t}
{m_{\mathrm{pbh}}^*}\approx 2\times 10^5\Msun \({t\over {1s}}\).
\q
The fraction of PBHs in all the dark matter, $\fpbh\equiv \Omega_{\mathrm{PBH}}/\Omega_{\mathrm{DM}}$ can be estimated by \cite{Nakama:2016gzw}
\m\label{fpbh}
f_{\mathrm{pbh}}\simeq 2.5 \times 10^{8}\beta\left(\frac{g_*^{\mathrm{form}}}{10.75}\right)^{-\frac{1}{4}}\left(\frac{m_{\mathrm{pbh}}}{M_{\odot}}\right)^{-\frac{1}{2}}.
\n
On comoving slices, the relation between the primordial comoving curvature perturbation, $\zeta(k)$, and the density contrast, $\Delta(k)$, at linear order is given by
\e\label{delta-zeta}
\Delta(k)=\frac{2(1+w)}{5+3w}\({k\over aH}\)^2\zeta(k),
\q
where $w$ is the equation of state and $H$ is the Hubble parameter. The comoving curvature perturbation $\zeta$ is related to the metric perturbations by 
\e
\zeta \equiv \psi - \mH(v+B),
\q
where our notations for the scalar perturbations $\psi$, $v$ and $B$ are introduced in Eq.~(\ref{gmunu}) and Eq.~(\ref{Tmunu}) below. Moreover, the comoving curvature perturbation is related to the Bardeen potential (see Eq.~(\ref{Phi}) below) by 
\m
\zeta = \Psi-\frac{2}{3(1+w)}(\mH^{-1} \Psi' +\Phi)
\n
For adiabatic perturbations, $\zeta$ stays constant on superhorizon scales. Then we have $\Phi=-3(1+w)/(5+3w)\zeta$ by assuming the absence of anisotropies and $\Phi=-2/3\zeta$ during RD. The density contrast smoothed over a scale, $R$, is calculated as
\e
\Delta(\bm{x},R)=\int \mathrm{d}^3 x' W(|\bm{x}-\bm{x'}|,R)\Delta(\bm{x'}),\qquad\Delta(\bm{k},R)=W(k,R)\Delta(k),
\q
with $W$ to be the window function chosen to smooth the density contrast. The variance of $\Delta(\bm{k},R)$ is given by
\e\label{squareroot}
\<\Delta^2\>=\int_{0}^{\infty}\frac{\mathrm{d}k}{k}W^2(k,R){ 4(1+w)^2\over (5+3w)^2}(kR)^4 P_\zeta(k),
\q
where we define the dimensionless curvature power spectrum as $\<\zeta(k)\zeta(k')\>\equiv \frac{2\pi^2}{k^3}\delta(\bm{k}+\bm{k'})P_\zeta(k)$. 
Usually, the amplitude of perturbations is assumed to obey Gaussian distribution. Once the perturbation satisfies the formation criterion, then PBHs are generated. Therefore, the formation of PBHs can be regarded as the statistics of peaks of a three-dimensional Gaussian random field, also known as peak theory \cite{Bardeen:1985tr}.
PBHs might be generated by very large perturbations with amplitudes $\mathcal{O}(0.01-0.1)$ if PBHs constitute most of the DM in our Universe. As a result, the primordial power spectrum that generate PBHs will have a ``bump'' on scales much smaller than the CMB scales \cite{Ivanov:1994pa,GarciaBellido:1996qt,Ivanov:1997ia,Yokoyama:1995ex,Kawasaki:2006zv,Hertzberg:2017dkh,Inomata:2017vxo,Inomata:2017okj,Kohri:2018qtx}.
As a benchmark example, we will assume that the power spectrum is ``spicky'' at a particular scale, $k_*$, to calculate $\beta$. In this case, the production of PBHs is almost monochromatic. Otherwise, for a broad power spectrum, one has take into account the so-called ``cloud-in-cloud'' problem where a single PBH is swallowed by the formation of a bigger PBH. Moreover, there are perturbations with different frequency and different amplitudes which may blur the calculation. 

For spicky spectrum, using peak theory, the number density of peaks can be approximated by (see e.g., Eq.~(4.14) in \cite{Bardeen:1985tr})
\e\label{peakt}
n_{\mathrm{pk}}(\nu_c)\simeq\frac{\(\<k^2\>/3\)^{3/2}}{(2\pi)^2}(\nu_c^2-1)\mathrm{e}^{-\nu_c^2/2},
\q
where the dimensionless threshold is defined as $\nu_c\equiv \Delta_c/\sqrt{\<\Delta^2\>}$. $\<k^2\>$ is defined as
\e
\<k^2\>=\frac{1}{\Delta^2}\int_{0}^{\infty}\frac{\mathrm{d}k}{k}k^2W^2(k,R)P_\zeta(k)
\q
Moreover, $\beta$ is related to $n_\mathrm{pk}$ by $\beta=n_\mathrm{pk}(\nu_c)(2\pi)^{3/2}R^3$.
Another commonly used method to calculate $\beta$ is the Press-Schechter formalism, where $\beta$ is evaluated by simply integrating the probability density function (PDF) beyond the threshold value
\e\label{PSbeta}
\beta=\int_{\nu_c}^{+\infty} \frac{\mathrm{d} \nu}{\sqrt{2 \pi}} e^{-\nu^{2} / 2 }=\half\mathrm{erfc}\({\nu_{c}\over\sqrt{2}}\).
\q
The Press-Schechter formalism only considers the amplitude of the perturbation but neglecting the higher derivatives. The comparison between Press-Schechter formalism and peak theory can be found in \cite{Young:2014ana} and the result showed close agreement differing by a factor of $10$ for large $\nu_{c}$.

%According to the simplicity of the Press-Schechter formalism, it is more widely used in literature to estimate the mass function of PBHs.
%For example, the Press-Schechter formalism is used to study the effects of primordial  non-Gaussianities (where the PDF of perturbations is non-Gaussian) \cite{Young:2013oia,Cai:2018dig,Cai:2019amo,Cai:2019elf,Yuan:2020iwf}.
The formation of PBHs depend on many aspects and there are some potential problems in calculating the mass function of PBHs. We will discuss the relevant aspects next.

\subsection{Primordial non-Gaussianities}
Formation of PBHs takes place at the tail of the PDF of the perturbations. Hence any non-Gaussianities that alter the PDF could significantly change the formation probability of PBHs. The impacts of non-Gaussianities on the formation of PBHs have been discussed long ago \cite{Bullock:1996at,Ivanov:1997ia,PinaAvelino:2005rm,Hidalgo:2007vk,Klimai:2012sf}. Besides considering a certain inflation model, a commonly used non-Gaussian model is the local-type non-Gaussianities, where the perturbation is expanded by the Gaussian part such that (up to cubic order)
\e\label{localNG}
\zeta=f(\zeta_g) = \zeta_g +\Fnl\(\zeta_g^2-\<\zeta_g^2\>\)+\Gnl \zeta_g^3.
\q
Here $\zeta_g$ is the Gaussian part whose PDF is Gaussian. The non-Gaussian parameter $\Fnl$ would skew the PDF and $\Gnl$ will change the kurtosis of the PDF.
Since the Press-Schechter formalism simply integrating the PDF above the threshold value, it is convenient in estimating the mass function $\beta$ in the presence of non-Gaussianities.
Let ${P_{\rm{NG}} }$ to be the PDF of $\zeta$ and ${P_{\rm{G}} }$ to be the PDF of the Gaussian part, $\zeta_g$.
Then ${P_{\rm{NG}} }$ can be obtained by changing the variables, $\zeta_g=f^{-1}(\zeta)$, such that
\m\label{changev}
\beta=\int_{\zeta_c}^{\infty}
{P_{\rm{NG}} }(\zeta)\mathrm{d}\zeta=
\int_{f_i^{-1}(\zeta)>\zeta_c}
\sum_{i=1}^{n}
{\mathrm{d}f_i^{-1}(\zeta)\over \mathrm{d}\zeta}P_\mathrm{G}[f_i^{-1}(\zeta)]\mathrm{d}\zeta.
\n
This is equivalent to integrate the Gaussian PDF in the region, $f_{i}^{-1}(\zeta)>\zeta_c$. Here, the lower index $i$ indicates the $i$-th solution of the total $n$ real solutions.
In \cite{Byrnes:2012yx}, Byrnes {\it et al}. used the above method to calculated $\beta$ within local-type non-Gaussianities up to cubic order. After that, the formation of PBHs within non-Gaussianities was studied up to fifth-order \cite{Young:2013oia} and their results showed a highly sensitive relation between $\beta$ and non-Gaussian parameters ($\Fnl$, $\Gnl$, etc.).

Except for changing the variables in Eq.~(\ref{changev}), \cite{Franciolini:2018vbk} adopted another method, the path-integral formulation, to calculate $\beta$ in non-Gaussian regions. The authors expressed $\beta$ by the sum of the N-point correlation function. Readers interested in the path-integral formulation can refer to their work and the references therein.

However, the path integral formulation might be impractical to calculate $\beta$. Riccardi {\it et al.} argued that the non-Gaussianities would affect cumulants at any order \cite{Riccardi:2021rlf}. Therefore, to get the exact result for $\beta$, one has to sum over all the N-point correlation function, which is impractical. They also proposed a semi-analytical expression to estimate the mass function of PBHs, see Eq.~(8) in \cite{Riccardi:2021rlf}.
%Besides local-type non-Gaussianities, another non-Gaussian model introduced in \cite{Nakama:2016kfq} whose PDF is described by
%\e
%P_{\rm{NG}}(\zeta)=\frac{1}{2\sqrt{2}\tilde{\sigma}\Gamma(1+1/p)}
%\exp\[-\({
%	|\zeta|
%	\over \sqrt{2}\tilde{\sigma}}\)^p
%\],
%\q
%where $\tilde{\sigma}$ and $p$ are positive parameters. This PDF will return to the Gaussian case if $p=2$. In \cite{Nakama:2016gzw}, the authors considered this non-Gaussian model and calculate $\beta$. 

\subsection{The non-linear effects between density contrast and curvature perturbation}
The standard procedure to calculate the density contrast is based on Eq.~(\ref{delta-zeta}). However, this equation is just the linear relation between the density contrast and the comoving curvature perturbation. In the comoving slicing, the non-linear relation in the long-wavelength approximation is given by \cite{Harada:2015yda,Yoo:2018kvb,Musco:2018rwt}
\m\label{nl}
\Delta(r,t)=-\frac{4(1+w)}{5+3w}\({1\over aH}\)^2\mathrm{e}^{-5\zeta(r)/2}\nabla^2\mathrm{e}^{\zeta(r)/2}.
\n
Due to the linear relation, the PDF of density contrast will no longer obey Gaussian distribution even if the PDF of curvature perturbations is Gaussian. This is an unavoidably generated non-Gaussianities that make the PBHs inevitably form in non-Gaussian regions. Recently, these ``intrinsic non-Gaussianities'' generated by the non-linear effects were studied in \cite{Yoo:2018kvb,Kawasaki:2019mbl,DeLuca:2019qsy,Young:2019yug} using either peak theory or Press-Schechter formalism. They found that the ``intrinsic non-Gaussianities would slightly suppress the PBH formation. More precisely, to produce the same abundance of PBH in our Universe, the power spectrum of $\zeta$ need to be amplified by a factor of $\sim\mathcal{O}(2)$ if using the linear relation, Eq.~(\ref{delta-zeta}), to calculate the abundance.

\subsection{Window Function and Power spectrum} 
From Eq.~(\ref{squareroot}), it is clear that the formation of PBHs depends on the choice of window functions and the power spectrum of the primordial scalar perturbations.
The power spectrum can be given by the inflation model, while the choice of the window function is a coarse-graining procedure.
Despite several commonly used window functions in literature, there is no physical interpretation on which window function should be used. The choice of window function will lead to uncertainties in calculating the mass function of PBHs, see \cite{Ando:2018qdb,Young:2019osy,Tokeshi:2020tjq}

The abundance of PBHs introduced in Eq.~(\ref{peakt}) and Eq.~(\ref{PSbeta}) can be applied to a narrow power spectrum, where the PBHs are about monochromatic and one can assume that the PBHs are formed at the same time. However, for a broad power spectrum, PBHs of different masses are expected to be formed at a different time, and the formation process will be rather complicated.

For the broad case, the Press-Schechter formalism failed, and one has to adopt the peak theory to evaluate the results. In \cite{MoradinezhadDizgah:2019wjf,DeLuca:2020ioi}, the authors studied the formation of PBHs for a tilted broad power spectrum, namely
\e
P_\zeta=A(k/k_s)^{n_p}\Theta(k_s-k)\Theta(k-k_l),
\q
where $\Theta$ denotes the Heaviside theta function. For a broad and flat spectrum, $n_p=0$, the corresponding mass distribution of PBHs is dominated by a single $\mpbh$. They also found that the mass function has a power-law tail, scaling as $\mpbh^{-3/2}$.

Recently, Yoo {\it et al.} \cite{Yoo:2020dkz} proposed a modified procedure to calculate PBH abundance in which they consider the intrinsic non-Gaussian effects within an arbitrary power spectrum in the absence of primordial non-Gaussianities. The results for a narrow spectrum showed no window function dependence, while the results for a broad power spectrum depends largely on the choice of window functions. They concluded that the top-hat window function in Fourier space would be the best choice since it minimizes the required property in theoretical PBH estimation.

\subsection{The formation criterion} Studies on the critical value of PBH formation can be traced back since 1975 when Carr first estimated the value of $\Delta_c$ by considering simplified Jeans length in Newtonian gravity, $\Delta_c=\cs$ \cite{Carr:1975qj}, where $c_s=1/\sqrt{3}$ is the sound speed during RD. However, the exact value of $\Delta_{c}$ is still a debating question since it depends on the initial density profile \cite{Polnarev:2006aa}, the equation of state and the sound speed \cite{Jedamzik:1996mr,Byrnes:2018clq}, primordial non-Gaussianities \cite{Kehagias:2019eil}, the primordial scalar power spectrum \cite{Germani:2018jgr} and even the window function \cite{Young:2019osy}.

Despite some spherically numeric simulations \cite{Jedamzik:1999am,Shibata:1999zs,Musco:2004ak,Hawke:2002rf,Polnarev:2006aa,Musco:2008hv,Musco:2012au,Nakama:2013ica,Harada:2015ewt,Harada:2015yda,Musco:2018rwt,Escriva:2019phb} on the formation of PBHs, Harada {\it et al.}  proposed an analytical expression, $\Delta_{c}=[3(1 + w)/(5 + 3w)] \sin^2[\pi\sqrt{w}/(1 + 3w)]\simeq0.41$ during RD \cite{Harada:2013epa}. Escriva {\it et al.} also proposed an approximate expression for $\Delta_c$ for $w\in[1/3,1]$ \cite{Escriva:2020tak}. The analytical $\Delta_c$ in a general cosmological background is studied by \cite{Escriva:2020tak} where they consider the equation of state to be $w\in(0,1]$.

Recently, \cite{Kehagias:2019eil} estimated the effects of ``intrinsic non-Gaussianities'' generated by the non linear relation, Eq.~(\ref{nl}), on $\Delta_{c}$. They found that the relative change of $\Delta_{c}$ is at the percent level, which might not be significant due to other uncertainties in estimating the formation of PBHs.
More recently, a detailed study based on numeric simulation was made by Musco {\it et al.} \cite{Musco:2020jjb} where they consider the intrinsic non-Gaussian effects and some commonly used power spectrum that generate the PBHs. They also gave an analytical approach to estimate $\Delta_c$ for all possible shapes of the power spectrum, see Eq.~(19) in \cite{Musco:2020jjb}.

Apart from using $\Delta_{c}$, Shibata and Sasaki proposed that the compaction function, which equals to half of the volume average of the density per  turbations in the long-wavelength limit, can be used to described the formation of PBHs \cite{Shibata:1999zs}. A general definition for the compaction function was introduced in \cite{Harada:2015yda} such that $\mathcal{C}=2\delta M/R$. Here $R$ represents the areal radius and $\delta M$ stands for the mass difference between the Misner-Sharp mass insides a sphere of radius $r$ and the mass inside a sphere with the same radius in the FLRW universe.

Using the maximum value of the compaction function, $\mathcal{C}_{\max}$, as a formation criterion, there would be less dependence on the shape of the density perturbation \cite{Shibata:1999zs} and this is later confirmed numerically by \cite{Harada:2015ewt}. \cite{Harada:2015ewt} also found that there would be less dependence on the lapse function if using $\mathcal{C}_{\max}$ instead of $\Delta_{c}$. Recently, it was shown in \cite{Escriva:2019phb} that the threshold value of compaction function, $\mathcal{C}_{\max,c}$, is only sensitive to the curvature at the maximum and $\mathcal{C}_{\max,c}$ is, to some approximations, a universal quantity to describe the PBH formation.

\section{Scalar-induced gravitational waves}
In this section, we will introduce the calculations of SIGWs during RD.
Let's begin from the most generic perturbed metric, which contains scalar modes and tensors modes, namely
\m\label{gmunu}
g_{00}&=&-1-2\phi,\no\\
g_{0i}&=&a\partial_iB,\no\\
g_{ij}&=&a^2\delta_{ij}+a^2 \(\half h_{ij}-2\delta_{ij}\psi+2\p_i\p_jE\),
\n
where $\phi$, $B$, $\psi$ and $E$ are linear scalar perturbations while the vector perturbations are not considered here. For vector-induced GWs and tensor-induced GWs, we refer the readers to \cite{Gong:2019mui}. Here, $h_{ij}$ denotes the second-order tensor mode. We do not consider the first-order tensor mode, since the second-order effects will far exceed the linear order effects during the formation of PBHs \cite{Saito:2008jc}. 
During RD, the cosmological background is described by perturbed perfect fluid, which reads
\m\label{Tmunu}
T_{00}&=&\rho+2\rho\phi+\delta\rho,\no\\
T_{0i}&=&-\rho\p_iB-P\p_i v-\rho\p_i v,\no\\
T_{ij}&=&(P+\delta P)\delta_{ij}-2P\psi\delta_{ij}+2P\p_i\p_jE,
\n
where $v$ is the velocity potential, $P$ and $\rho$ are  the pressure and the energy density of the background while $\delta P$ and $\delta \rho$ are the corresponding first-order perturbations. Here we neglect the anisotropies caused by neutrinos and photon \cite{Saga:2014jca}. 

As we shall see, besides the GWs, there is only one physical degree of freedom in our calculation. Consider a first-order change in the coordinate such that
\e
\tilde{\eta}= \eta+T,~\tilde{x}^i= x^i+\p^iL,
\q
and the scalar modes will transform as \cite{Malik:1998ai}
\m\label{1st-GT}
\tilde{\phi}&=&\phi+\mH T+T',\\
\tilde{\psi}&=&\psi-\mH T,\\
\tilde{B}&=&B-T+L',\\
\tilde{E}&=&E+L,
\n
where a prime denotes a derivative with respect to the conformal time $\eta$. The most generic perturbed metric has four scalar modes. However, one can fix $T$ to determine the time-slicing and choose $L$ to decide the spatial coordinates on the hyper-surfaces. This would reduce two degrees of freedom. Furthermore, the Einstein equation will reduce one more degree of freedom and give the equation of motion for the last degree of freedom. 

The simplest gauge to calculate the SIGWs is the Newton gauge, also known widely as orthogonal zero-shear gauge, longitudinal gauge. Newton gauge demands that $\tilde{B}=\tilde{E}=0$. This can be done by choosing $L=-E$ and $T= B-E'$. Then, the remaining two modes are just the Bardeen potential
\m\label{Phi}
\Phi\equiv\phi+\mH\sigma+\sigma',\quad\Psi\equiv\psi-\mH\sigma,
\n
where $\sigma\equiv E'-B$ is the shear potential. The advantage of the Newton gauge is that the degrees of freedom are completely fixed, and there are no more gauge modes. The only modes left are just the Bardeen potential $\phi=\Phi$ and $\psi=\Psi$. Write down the first-order Einstein equation in terms of the Bardeen potential, one finds that $\Phi=\Psi$ and the equation of motion for the only scalar mode left is
\e\label{phi1}
\Phi''+4\mH \Phi-{1\over 3}\nabla^2\Phi=0.
\q
Keeping the decay modes of Eq.~(\ref{phi1}), the solution of $\Phi$ in Fourier space is given by 
\m\label{phi_sol}
\Phi(\vec{k},\eta)\equiv\Phi_k
T_\Phi(k\eta)=\Phi_k{9\over x^2}\({\sin(x/\sqrt{3})\over x/\sqrt{3}}-\cos(x/\sqrt{3})\),
\n
where we label the primordial value as $\Phi_k$ and $T_\Phi$ is the normalized transfer function such that $T_\Phi(0)=1$. We also introduce the dimensionless variable $x\equiv k\eta$. The value of $\Phi_k$ is given by certain inflation models and we will treat it as free parameters in our discussion.
Moreover, the first-order GWs is given by 
\e\label{GW1}
h_{i j}^{\prime \prime(1)}+2 \mathcal{H} h_{i j}^{\prime(1)}-\nabla^{2} h_{i j}^{(1)}=0,
\q
which is a source-free equation. The linear GWs will decay from the primordial value as $\eta^{-1}$ during RD. During the formation of PBHs, the scalar power spectrum are enhanced to $\mathcal{O}(0.01)$ if PBH represents the main part of the DM. In this case, the second-order SIGWs would be stronger than the linear GWs, and hence we will not consider $h_{ij}^{(1)}$. Readers interested in second-order GWs induced by $h_{ij}^{(1)}$ can refer to \cite{Gong:2019mui}.

Up to second-order, the evolution of the GWs can be written as
\e\label{eqh}
h_{i j}^{\prime \prime}+2 \mathcal{H} h_{i j}^{\prime}-\nabla^{2} h_{i j}=-4 \mathcal{T}_{i j}^{\ell m} S_{\ell m},
\q
where $\mathcal{T}_{i j}^{\ell m} = e_{ij}^{(+)}(\bm{k})e^{(+)lm}(\bm{k})+e_{ij}^{(\times)}(\bm{k})e^{(\times)lm}(\bm{k}) $ is the transverse and traceless projection operator. The polarization tensors are defined as $(e_i e_j - \bar{e}_i \bar{e}_j)/\sqrt{2}$ and $(e_i \bar{e}_j + \bar{e}_i e_j)/\sqrt{2}$ for $+$ and $\times$ modes. Here we choose $e=(1,0,0)$, $\bar{e}=(0,1,0)$ and $\bm{k}=(0,0,k)$.

The source term in eq.~(\ref{eqh}) reads
\m
S_{ij}=3\Phi\p_i\p_j\Phi-{2\over\mH}\p_i\Phi'\p_j\Phi-{1\over\mH^2}\p_i\Phi'\p_j\Phi',
\n
where we have expressed the density perturbation $\delta\rho$, the velocity potential $v$ and the pressure perturbation $\delta P$ in terms of the scalar modes $\Phi$. 
Overall, the perfect fluid acts like an intermediary, and the GWs seem to be generated by the scalar modes. This is why SIGWs gain its name.
Eq.~(\ref{eqh}) can be solved by Green's function in Fourier space, namely
\e\label{solh}
h(\eta,\vec{k})={1\over a(\eta)}\int_0^\eta g_k(\eta;\eta')a(\eta')S(\eta',\vec{k})\mathrm{d}\eta'.
\q
We define $S(\eta,\vec{k})\equiv -4e^{ij}(\vec{k})\mathcal{S}(\eta,k)$ and $\mathcal{S}(\eta,k)$ is the source term transferred to Fourier space. Here, we define the Fourier transform of $h_{ij}$ to be
\e\label{h_ft}
h_{ij}(\eta,\bm{x})=\int{\mathrm{d}^3k \over (2\pi)^3}\[ h^{(+)}(\eta,\bm{k})e_{ij}^{(+)}(\bm{k}) + 
h^{(\times)}(\eta,\bm{k})e_{ij}^{(\times)}(\bm{k}) \]e^{i \bm{k}\cdot\bm{x}}.
\q
In the following part, we write $h(\eta,\bm{k})$ to denote either the plus mode or the cross mode.
In our convention, $S(\eta,\vec{k})$ takes the form
\m
S(\eta,\vec{k})=-4\int\frac{\mathrm{d}^3p}{(2\pi)^{3/2}} \Big(e^{ij}p_ip_j\Big)\Phi_p\Phi_{|\vec{p}-\vec{k}|}F(|\vec{p}|,|\vec{k}-\vec{p}|,\eta).
\n
The transfer function is defined as
\m
{F}(u,v,x)=3T_\phi(ux)T_\phi(vx)+uxT_\phi'(ux)T_\phi(vx)+vxT_\phi'(vx)T_\phi(vx)+uvx^2T_\phi'(ux)T_\phi'(vx).
\n
For convenience, we introduce the dimensionless variable $u\equiv p/k$, $v\equiv |\vec{p}-\vec{k}|/k$ and $x\equiv k\eta$. Unless otherwise being stated, the prime with $T'(y)$ denotes the derivative with respect to $y$, other than the conformal time.
The Green's function in Eq.~(\ref{solh}) takes the form $g_k(\eta;\eta')={1\over k}\sin(k\eta-k\eta')$ during RD. A more important quantity in observation is the density parameter of the stochastic GW background defined as the energy of GWs per logarithm frequency normalized by the critical energy $\rho_{c}(\eta)$
\e\label{ogwd}
\ogw(k,\eta)\equiv{1 \over \rho_c}{\mathrm{d}\rho_{\mathrm{GW}} \over \mathrm{d} \ln k}
={k^3\over48\pi^2}\(k\over\mH\)^2\overline{\<|h(\eta,{\vec{k}})|^2\>},
\q
where the overline denotes the oscillating average.
To calculate the stochastic GW background of the SIGWs, we have to know the primordial scalar power spectrum, which is given by specific inflation models. In this paper, we do not consider certain inflation model, instead, we will parameterize the power spectrum by some commonly used function in literature.
The density parameter is evaluated as (see \eg \cite{Kohri:2018awv})
\m\label{ogw(k)}
\ogw(k,\eta)\!\!\!\!\!\!\!\!\!\!\!&&= \(\frac{k}{\mH}\)^2{k^3\over48\pi^2{a(\eta)}^2}\int\mathrm{d}\tilde{\eta_1}\mathrm{d}\tilde{\eta_2}{{g_k(\eta;\tilde{\eta_1})g_{k'}(\eta;\tilde{\eta_2})}}
a({\tilde{\eta_1}})a(\tilde{\eta_2}){\<{S}(\tilde{\eta_1},\vec{k}){S}(\tilde{\eta_2},\vec{k'})\>}.\no\\
%%%%%%%%%%%%%%%%%%%%%%%%%%%%%%%%%
&&=\(\frac{k}{\mH}\)^2{4\pi^2k^3\over3a(\eta)^2}\int\frac{\mathrm{d}^3p}{(2\pi)^3} 
\int\mathrm{d}\tilde{\eta_1}\mathrm{d}\tilde{\eta_2}{a(\tilde{\eta_1})a(\tilde{\eta_2})}{g_k(\eta;\tilde{\eta_1})g_k(\eta;\tilde{\eta_2})}
\Big(e^{ij}p_ip_j\Big)^2{1\over p^3|\vec{k}-\vec{p}|^3}\no\\
&&\times P_\Phi(k)P_\Phi(|\vec{k}-\vec{p}|)\Big[{F}(|\vec{p}|,|\vec{k}-\vec{p}|,\tilde{\eta_1}){F}(|\vec{p}|,|\vec{k}-\vec{p}|,\tilde{\eta_2})+{F}(|\vec{p}|,|\vec{k}-\vec{p}|,\tilde{\eta_1}){F}(|\vec{k}-\vec{p}|,|\vec{p}|,\tilde{\eta_2})\Big]\no\\
%%%%%%%%%%%%%%%%%%%%%%%%%%%%%%%%%%%%%%%%%
&&=\(\frac{k}{\mH}\)^2{8\pi^2k\over3}\int\frac{\mathrm{d}^3p}{(2\pi)^3}\(\int_0^{\eta}\mathrm{d}\tilde{\eta_1}{a(\tilde{\eta_1})\over a(\eta)}kg_k(\eta;\tilde{\eta_1})\tilde{F}(|\vec{p}|,|\vec{k}-\vec{p}|,\tilde{\eta_1})\)^2
\Big(e^{ij}p_ip_j\Big)^2{1\over p^3|\vec{k}-\vec{p}|^3}\no\\
&&\qquad\times P_\Phi(k)P_\Phi(|\vec{k}-\vec{p}|)\no\\
\label{gw}
&&\simeq{1\over 6}\int_0^\infty\mathrm{d}u\int_{|1-u|}^{1+u}\mathrm{d}v~{v^2\over u^2}\Big[1-\({1+v^2-u^2\over 2v}\)^2\Big]^2P_\Phi(uk)P_\Phi(vk)\overline{I^2(u,v,x)},
\n
where in the last step of Eq.~(\ref{gw}), we define the power spectrum as
\m
\<\Phi(\vec{k})\Phi(\vec{k'})\>\equiv {2\pi^2\over k^3}P_\Phi(k)\delta(\vec{k}+\vec{k'}),
\n
and the kernel function to be
\m\label{is}
I(u,v,x)\equiv\int_0^x\mathrm{d}\tx~\tx\sin(x-\tx)\tilde{F}(u,v,x),
\n
where $\tilde{F}(u,v,x)\equiv (F(u,v,x)+F(v,u,x))/2$ is the symmetric part of the transfer function. We also used that $\mH=\eta^{-1}$ and $a(\tilde{\eta})/a(\eta)=\tilde{\eta}/\eta$ during RD. We sum over the two polarization modes in Eq.~(\ref{ogw(k)}) and the $e^{ij}$ in the expression refers to either the ``$+$'' mode or the ``$\times$'' mode. In the last step of Eq.~(\ref{ogw(k)}), the overline denotes the oscillating average, namely $\sin^2 x\to \half$, $\cos^2 x\to\half$ and $\sin x\cos x\to0$. The $\Omega_{\mathrm{GW}}$ has a time dependence through $I(u,v,x)$ and, as we will demonstrate below, the function $I(u,v,x)$ will converge to a finite value at late time.

The density parameter given by Eq.~(\ref{ogw(k)}) is independent of the position, rendering the SIGWs are isotropic GWs. This is due to the statistical homogeneity of the FLRW metric. However, the GW detectors would detect GWs from different directions, resulting in angular anisotropies of the SIGW signal. This was studied in \cite{Bartolo:2019zvb}. In the absence of primordial non-Gaussianities, the authors found that the anisotropies are negligible by today due to the propagation effects. 

After the horizon entry, the energy of GWs redshift as radiation,  $\rho_{\rm{GW}}\propto a^{-4}$. Therefore, the current GW density parameter, $\Omega_{\mathrm{GW,0}}$ would be 
\e
\Omega_{\mathrm{GW,0}}(\eta_0,f) = \Omega_r \Omega_{\mathrm{GW}}(\eta_c,f),
\q
where $\Omega_r$ is the energy density fraction of radiation by today and we neglect the effects of relativistic degrees of freedom. Here, $\Omega_{\mathrm{GW}}(\eta_c,f)$ is the late time value when $\Omega_{\mathrm{GW}}(\eta,f)$ becomes a constant and can be computed by taking $I(u,v,x\to\infty)$ (Here, a constant means that $\Omega_{\mathrm{GW}}(\eta,f)$ is independent of time).
This result would be more clear if one considers the evolution of tensor modes during both MD and RD. See e.g., \cite{Kohri:2018awv} where the author studied analytically the SIGWs during MD, RD and RD-to-MD transition.
The analytical expression for $\overline{I^2(u,v,x)}$ was derived in \cite{Espinosa:2018eve,Kohri:2018awv}. The indefinite integral of Eq.~(\ref{is}) is given by
\m
I(u,v,x)\!\!\!\!\!\!\!\!\!\!&&=-{27(u^2+v^2-3)^2\over 16u^3v^3}
\Bigg\{
\Bigg[
\Si\(\(1-{(u+v)\over\sqrt{3}}\)x\)
+
\Si\(\(1+{(u+v)\over\sqrt{3}}\)x\)
-
\Si\(\(1+{(u-v)\over\sqrt{3}}\)x\)\no\\
&&
-\Si\(\(1-{(u-v)\over\sqrt{3}}\)x\)
\Bigg] \cos x
+\Bigg[
\Ci\(\(1+{(u-v)\over\sqrt{3}}\)x\)
+
\Ci\(\(1-{(u-v)\over\sqrt{3}}\)x\)\no\\
&&
-
\Ci\(\Bigg|1-{(u+v)\over\sqrt{3}}\Bigg|x\)
-
\Ci\(\(1+{(u+v)\over\sqrt{3}}\)x\)
\Bigg]\sin x
+{1\over u^2+v^2-3}\no\\
&&\times \(4uv+(u^2+v^2-3)\ln\Bigg|
1-\frac{4uv}{(u+v)^2-3}
\Bigg|\)\sin x+{12\over x^2(u^2+v^2-3)^2}
\no\\
&&\times \Bigg[
2u\cos{ux\over\sqrt{3}}\(vx\cos{vx\over\sqrt{3}}-\sqrt{3}\sin{vx\over\sqrt{3}}\)+2u\(-vx+\sqrt{3}\sin{vx\over\sqrt{3}}\)\no\\
&&+\sin{ux\over\sqrt{3}}\(
2\sqrt{3}v-2\sqrt{3}v\cos{vx\over\sqrt{3}}+(u^2+v^2-3)x\sin{vx\over\sqrt{3}}
\)
\Bigg]
\Bigg\}.
\n
Then the late time expression, $\overline{I^2(u,v,x\to\infty)}$, takes the form
\m\label{I2S}
\overline{I^2}=\frac{729(u^2+v^2-3)^2}{512u^6v^6}\Bigg\{\Big(-4uv+(u^2+v^2-3)\ln\Big|{3-(u+v)^2\over 3-(u-v)^2}\Big|\Big)^2+\pi^2\(u^2+v^2-3\)^2\Theta(u+v-\sqrt{3})\Bigg\}.
\n

Below we use the comoving curvature perturbation $\zeta$, which has the relation $\zeta=(3/2)\Phi$ to calculate the SIGWs. A widely used model for the power spectrum is the infinite narrow spectrum
\e\label{delta}
P_\zeta(k)=Ak_*\delta(k-k_*),
\q
with a dimensionless amplitude $A$ peaked at $k_*$. This power spectrum corresponds to a monochromatic PBH formation and is convenient to study the properties of SIGWs from PBHs. For this spectrum, the density parameter has an analytical form, namely
\e
\ogw(k)={3 A^2\Omega_{\rm{r}}\over 64}\tilde{k}^2\(1-{\tilde{k}^2\over4}\)^2(2-3\tilde{k}^2)^2\Theta(2-\tilde{k})\Bigg((2-3\tilde{k}^2)^2\pi^2\Theta(2-\sqrt{3}\tilde{k})+\(-4+(2-3\tilde{k}^2)\ln\Big|1-{4\over 3\tilde{k}^2}\Big|\)^2\Bigg),
\q
where we define $\tilde{k}\equiv k/k_*$.
Another example is a log-normal power spectrum parameterized by
\e\label{ln}
P_\zeta(k)={A\over \sqrt{2\pi\sigma_*^2}}\exp\(-\frac{\ln\tilde{k}^2}{2\sigma_*^2}\),
\q
where $\sigma_*$ is a dimensionless parameter which denotes the width of the power spectrum. The SIGWs generated by this power spectrum has been carefully studied in \cite{Pi:2020otn}.

\begin{figure}[htbp!]
	\centering
	\includegraphics[width = 0.6\textwidth]{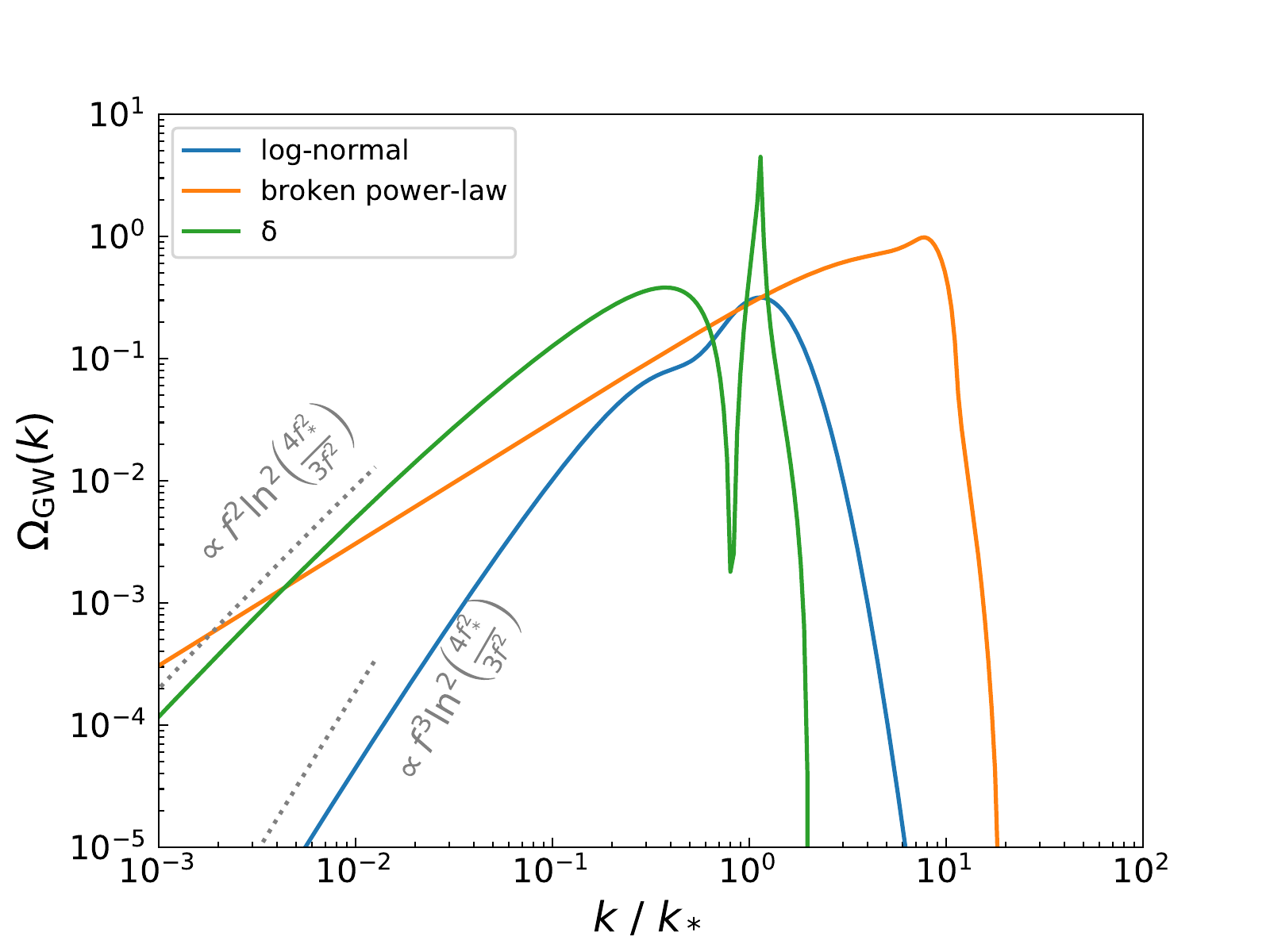}
	\caption{\label{ogw} SIGWs generated by different power spectrum. The parameters for the log-normal spectrum is $\sigma_*=0.5$. For the box spectrum, we set $k_{\mathrm{min}}=0.1$ and $k_{\mathrm{max}}=3$.
	}
\end{figure}

Finally, we consider a box spectrum described by 
\e\label{box}
P_\zeta(k)=\frac{A}{\ln\({k_{\mathrm{max}}\over k_{\mathrm{min}}}\)} \Theta(\tilde{k}-k_{\mathrm{min}})\Theta(k_{\mathrm{max}}-\tilde{k}).
\q
In the above case, we have normalized the power spectrum so that $\int P_\zeta(k)~\mathrm{d}\ln k = A = \<\zeta^2\> $ is the variance of the perturbations.
The $\Omega_{\mathrm{GW}}(k)$ for these three power spectrum is given in Fig.~\ref{ogw}.

\begin{figure}[htbp!]
	\centering
	\includegraphics[width = 0.6\textwidth]{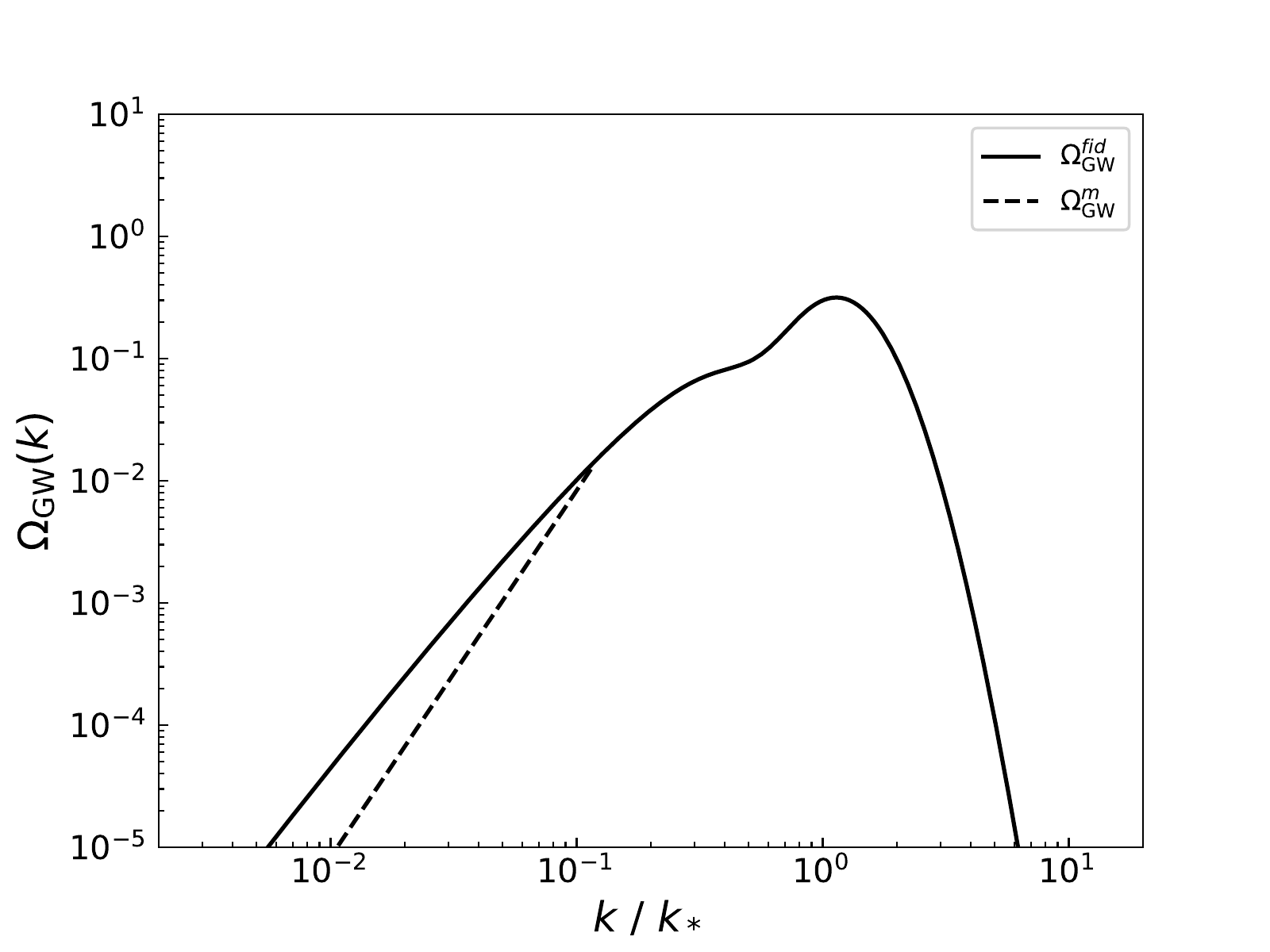}
	\caption{\label{case} The two models described by Eq.~(\ref{twocase}). Both lines correspond to the SIGWs induced by a log-normal power spectrum with $\sigma_*=0.5$. The solid line is the full numeric result, while the dashed line has a $k^3$ scaling in the infrared region, $k<0.1k_*$.
	}
\end{figure}

\begin{figure}[htbp!]
	\centering
	\includegraphics[width = 0.6\textwidth]{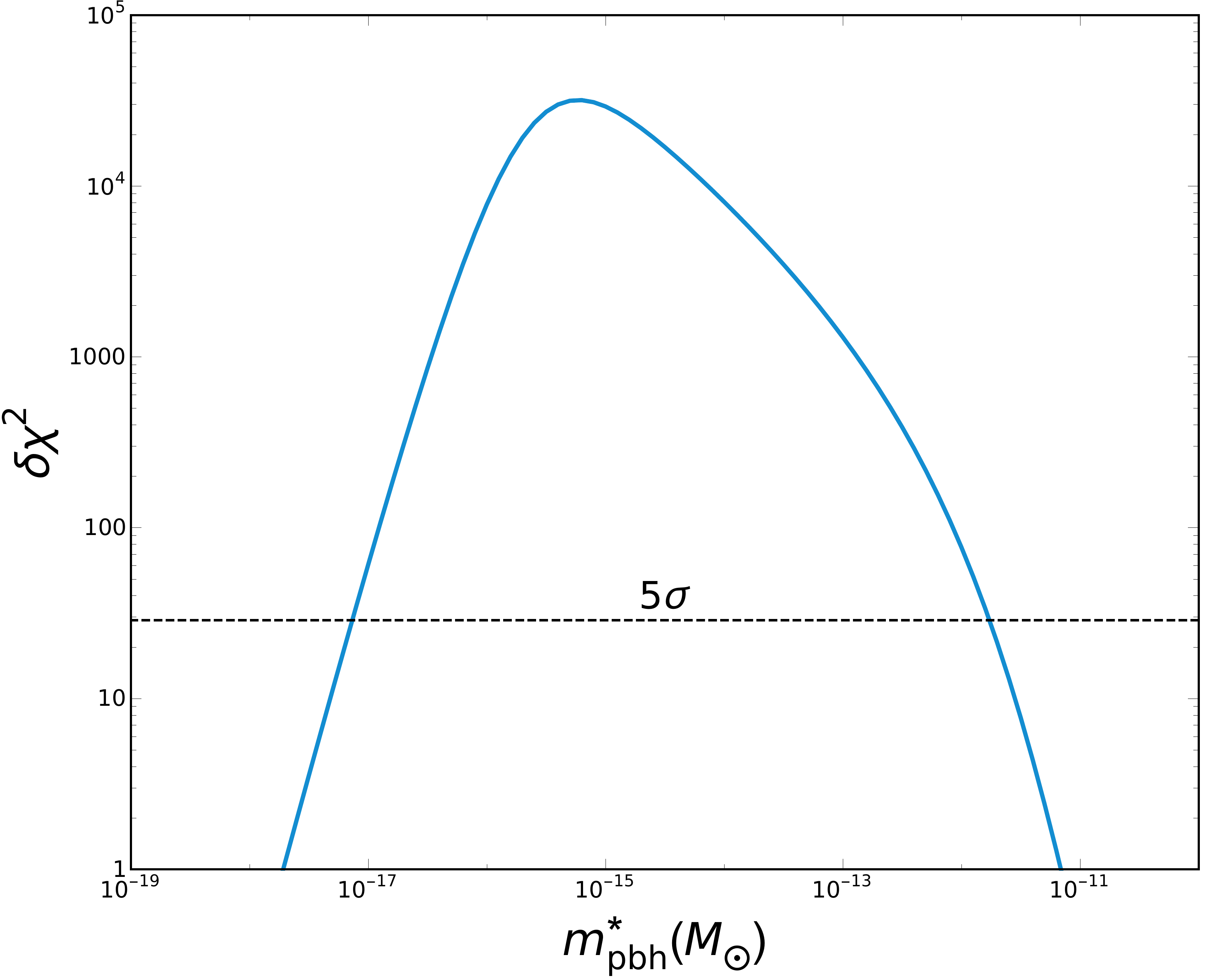}
	\caption{\label{chi} Taken from Fig.~3 in  \cite{Yuan:2019wwo}. The relation between $\delta\chi^2$ and the peak mass of PBHs ($\mpbh^*$ corresponds to $k_*$) generated by the log-normal power spectrum given in Eq.~(\ref{ln}) with $\sigma_*=0.5$. The amplitude of ${A}$ is fixed by assuming PBHs represent $10^{-3}$ of DM. The $5\sigma$ dashed line corresponds to $\delta\chi^2=28.74$. 
	}
\end{figure}

As shown in the figure, on small scales, the SIGWs will have a cut-off wavelength. The cut-off is due to the momentum conservation such that two modes $\bf{k_1}$ and $\bf{k_2}$ generate a mode with $\bf{k}=\bf{k_1}+\bf{k_2}$. Therefore, for a power spectrum defined on $[k_{\min},k_{\max}]$, the cut-off wavelength of the corresponding SIGWs will be $2k_{\max}$.
Since a constant gravitational potential will not change the distribution of matter, it will not induce GWs. After the perturbation re-enters the horizon at $k\simeq k_*$, they will induce GWs, and the shape of $\Omega_{\mathrm{GW}}$ is dominated by the power spectrum. For instance, a power spectrum with a peak located at $k_*$ will induce GWs with a peak at $2k_*/\sqrt{3}$ where $1/\sqrt{3}$ comes from the sound of speed during RD while a box spectrum will induce a smooth signal. Finally, on large scales, the scaling of SIGWs shows a log-dependent behavior, $\Omega_\mathrm{GW}\propto f^2\ln\({4f_*^2\over 3f^2}\)$ or $\Omega_\mathrm{GW}\propto f^3\ln\({4f_*^2\over 3f^2}\)$ where $f_*$ is a pivot scale of the power spectrum. On large scales, the power spectrum has decreased to negligible value, and the signal is dominated by the evolution of scalar perturbations. A strict proof of the log-dependent scaling is given in \cite{Yuan:2019wwo} for a general spectrum. This special scaling is very important in distinguishing the signal of SIGWs from other stochastic GW background, and it could be smoking guns in detecting SIGWs.
In \cite{Yuan:2019wwo}, Yuan {\it et al.} estimated the distinguishability of LISA. They consider a fiducial case, $\Omega_{\mathrm{GW}}^{fid}(k)$, generated by a log-normal power spectrum with $\sigma_*=0.5$ and the other one is described by 
\begin{equation}\label{twocase}
	\ogw^{m}(k)=\left\{
	\begin{aligned}
		&\ogw^{fid}(0.1k_*){\({k\over 0.1k_*}\)^3}~,&&\ \hbox{for}\ k<0.1k_*, \\
		&\ogw^{fid}(k)~,&&\ \hbox{for}\ k\ge0.1k_*. \\
		%z&&\frac xy
	\end{aligned}
	\right.
\end{equation}
These two cases have different scalings in the infrared region, as shown in Fig.~\ref{case}. 
A statistic quantity, $\delta\chi^2$, which characterize the discrepancy of two models, takes the form \cite{Kuroyanagi:2018csn}
\m
\delta\chi^2\simeq T\int_0^\infty\mathrm{d}f\(\frac{\ogw^{fid}-\ogw^{m}}{\ogw^{m}+\Omega_{n}}\)^2,
\n
where $T$ is the observation time and $\Omega_{n}$ is the noise density parameter of the detector.
Scanning the mass of PBHs, the result of $\delta\chi^2$ is Fig.~\ref{chi}.
It is shown that LISA can well distinguish the scaling of two models beyond $5\sigma$ in a wide mass range, especially for those PBHs of $\mpbh \in[10^{-16},10^{-14}] \cup [10^{-13},10^{-12}] M_\odot$ which could still represent all the DM in our Universe. To summarize, the log-dependent scaling, $\Omega_\mathrm{GW}\propto f^2\ln\({4f_*^2\over 3f^2}\)$ or $\Omega_\mathrm{GW}\propto f^3\ln\({4f_*^2\over 3f^2}\)$ can be smoking guns in detecting SIGWs.

Since there are no observational results concerning the primordial scalar power spectrum on small scales, the power spectrum is usually parameterized by some given function in literature. A more realistic way is to consider the primordial power spectrum from a given inflation model and then calculate the corresponding SIGWs. See reflevant works \cite{Kawasaki:2013xsa,Choudhury:2013woa,Inomata:2016rbd,Gong:2017qlj,Ando:2017veq,Ando:2018nge,Xu:2019bdp,Ozsoy:2019lyy,Gao:2019kto,Lin:2020goi,Ballesteros:2020qam,Liu:2020oqe,Braglia:2020eai,Fu:2020lob,Dalianis:2020cla,Yi:2020kmq,Dalianis:2020qjs,Aldabergenov:2020yok,Ragavendra:2020sop,Bhaumik:2020dor,Zhou:2020kkf,Ragavendra:2020vud,Yi:2020cut,Braglia:2020taf,Gao:2020tsa,Gao:2021vxb}. 

\subsection{Scalar Induced Gravitational Waves in a General Cosmological Background}
For a constant equation of state and let $\cs=w$, the semi-analytical solutions of SIGWs were obtained by Dom\`enech in \cite{Domenech:2019quo} for $0<w\le1$. Then, the results for constant $w$ were extended to $w<0$ in \cite{Domenech:2020kqm}. For the most general situation where $w$ and $\cs$ may varied, there are no semi-analytical solutions to SIGWs, and a numeric method should be adopted. For instance, during the QCD phase transition, $w$ and $\cs$ may become slightly smaller than $1/3$ and the equation $\cs=w$ no longer holds. The SIGWs generated during the QCD epoch were numerically calculated in \cite{Abe:2020sqb}.
%For general $w(\eta)$ and $c_s(\eta)$, the results are independent of the time derivatives of $w(\eta)$ and $c_s(\eta)$. This is because the Einstein equation gives $G_{\mu\nu}=\kappa T_{\mu\nu}$ which is irrelevant to derivatives of the pressure and the density of the perfect fluid.
The derivation within a general $w(\eta)$ and $c_s(\eta)$ differs from that in RD in the following aspects.

Firstly, the equation of motion for linear scalar perturbations become
\e
\Phi''+3\mH(\eta)(1+\cs)\Phi'+3\mH^2(\cs-w)\Phi-\cs\nabla^2\Phi=0,
\q
where the conformal Hubble parameter now become $\mH(\eta)={2\over(1+3w)\eta}$. This equation should be numerically solved with the initial condition to be $\Phi_{\bm{k}}(\eta\to0)=\Phi_{k}T_\Phi(k\eta\to0)=\Phi_k$. 

Secondly, the equation of motion for the second-order tensor modes takes the same form as Eq.~(\ref{eqh}) but the source term changes to
\m
S_{ij}=2\Phi\p_i\p_j\Phi
-{4\over3(1+w)}
\(\p_i\Phi+{\p_i\Phi'\over\mH(\eta)}\)
\(\p_j\Phi+{\p_j\Phi'\over\mH(\eta)}\).
\n
The Green's function and the transfer function methods can still be applied to solve $h_{ij}$. Since the form of the equation of motion for $h_{ij}$ is unchanged, the solution to $h_{ij}$ takes the same form as Eq.~(\ref{solh}) except for the Green's function, which now satisfies the equation
\m
\(\p_{\eta}^2+k^2-{1-3w(\eta)\over2}\mH^2\)g_k(\eta,\tilde{\eta})=\delta(\eta-\tilde{\eta}),
\n
The solution of $g_k(\eta,\tilde{\eta})$ can be obtained by
\e
g_k(\eta,\tilde{\eta})=
\frac
{u(\eta)v(\tilde{\eta})-u(\tilde{\eta})v(\eta)}
{u'(\tilde{\eta})v(\tilde{\eta})-u(\tilde{\eta})v'(\tilde{\eta})},
\q
with $u(\eta)$ and $v(\eta)$ to be the two independent homogeneous solutions.
Move on to the density parameter, the final expression for $\Omega_{\mathrm{GW}}$ takes the same form as the last line of Eq.~(\ref{ogw(k)}) but the kernel function should now be evaluated as
\e
I(u,v,x)=k^2\int_0^\eta\mathrm{d}\tilde{\eta}{a(\tilde{\eta})\over a(\eta)}g_k(\eta,\tilde{\eta})\tilde{F}(|\bm{p}|,|\bm{k}-\bm{p}|,\tilde{\eta}).
\q
After calculating the oscillating average of $I(u,v,x)$, the density parameter, $\Omega_{\mathrm{GW}}$, can be obtained. Readers interested in numerically solutions may refer to \cite{Abe:2020sqb} for details where the authors introduce some technique to reduce the computational complexity. The $\Omega_{\mathrm{GW}}$ of SIGWs generated during the QCD phase transition is shown in Fig.~4 in \cite{Abe:2020sqb} for a delta-power spectrum.

\subsection{Higher-Order Corrections to Scalar Induced Gravitational Waves }
SIGWs are generated at second-order soured by linear scalar perturbations, which lead to $h_{ij}\sim \Phi^2$. Given that the scalar perturbations are enhanced to $\sim\mathcal{O}(0.01-0.1)$ during the formation of PBHs, the higher-order corrections to SIGWs are expected to be significant. The higher-order corrections to SIGWs were first calculated by Yuan {\it et al.} where they proposed a semi-analytical method to evaluate the SIGWs generated by a $\delta$-spectrum during RD \cite{Yuan:2019udt}.
Yuan {\it et al.} computed the tensor modes sourced by the quadratic, cubic and biquadratic terms of the linear perturbations during RD.
Here we give the source term in Newton gauge up to fourth-order in general cosmological background. Our convention for the higher-order tensor modes are (up to fourth-order)
\e
\delta g_{ij}=a^2\((1-2\Phi)\delta_{ij}+\half(h_{ij}^{(2)}+h_{ij}^{(3)}+h_{ij}^{(4)})\),
\q
and the perturbed perfect fluid up to third-order in Newton gauge is given by
\m
T_{00}^{(2)}\!\!\!\!\!&=&\!\!\!\!\!\half
\delta\rho^{(2)}+2\delta\rho^{(1)}\Phi^{(1)}+\rho\Phi^{(2)}+(P+\rho)v_k^{(1)}v^{k(1)}
 \no\\
T_{0i}^{(2)}\!\!\!\!\!&=&\!\!\!\!\!\half\Big(
-\rho B_i^{(2)}-2\delta\rho^{(1)}B_i^{(1)}
-2(\delta P^{(1)}+\delta\rho^{(1)}) v_i^{(1)} -(P+\rho) \(4h_{ik}^{(1)}v^{k(1)}+2\(\Phi^{(1)}-2\Psi^{(1)}\)v_i^{(1)}+v_i^{(2)}
\)
\Big)
\no\\
T_{ij}^{(2)}\!\!\!\!\!&=&\!\!\!\!\!Ph_{ij}^{(2)}+2h_{ij}^{(1)}\delta P^{(1)}+
(P+\rho)\(B_i^{(1)}+v_i^{(1)}\)\(B_j^{(1)}+v_j^{(1)}\)+
\half\delta_{ij}\(
\delta P^{(2)}-4\delta P^{(1)}\Psi^{(1)}-2P\Psi^{(2)}
\)\no\\
%%%%%%%%%%%%%%%%%%%%%%%%%%%%%%
T_{00}^{(3)}\!\!\!\!\!&=&\!\!\!\!\!
\(\delta P^{(1)}+\delta\rho^{(1)}\)v_k^{(1)}v^{k(1)}+(P+\rho)v^{k(1)}v_k^{(2)}
+2(P+\rho)h_{lm}^{(1)}v^{l(1)}v^{m(1)}+
{1\over 6}\delta\rho^{(3)}+{1\over3}\rho\Phi^{(3)}\no\\
&&+2(P+\rho)\Phi^{(1)}v_k^{(1)}v^{k(1)}
+
\delta\rho^{(2)}\Phi^{(1)}+\delta\rho^{(1)}\Phi^{(2)}-2(P+\rho)\Psi^{(1)}v_k^{(1)}v^{k(1)}
\no\\
%%%%%%%%%%%%%%%%%%%%%%%%%%%%%%
T_{0i}^{(3)}\!\!\!\!\!&=&\!\!\!\!\!{1\over 6}\Bigg[
(P+\rho)\Big[-6B_i^{(1)}B^{k(1)}v_k^{(1)}-6\(h_{ik}^{(2)}v^{k(1)}+h_{ik}^{(1)}v^{k(2)}\)+v_i^{(3)}-3\Big((2B_i^{(1)}+v_i^{(1)})v_k^{(1)}v^{k(1)}\no\\
&&+4h_{ik}^{(1)}v^{k(1)}\Phi^{(1)}+v_i^{(2)}(\Phi^{(1)}-2\Psi^{(1)})-v_i^{(1)}\((\Phi^{(1)})^2-\Phi^{(2)}+4\Phi^{(1)}\Psi^{(1)}+2\Psi^{(2)}\)
\Big)
\Big]
-3\(\delta P^{(1)}+\delta\rho^{(1)}\)\no\\
&&\times
\(4h_{ik}^{(1)}v^{k(1)}+v_i^{(2)}+2v_i^{(1)}(\Phi^{(1)}-2\Psi^{(1)})\)-3\(\delta P^{(2)}+\delta\rho^{(2)} \)v_i^{(1)}-3\(B_i^{(2)}\delta\rho^{(1)}+B_i^{(1)}\delta\rho^{(2)}\)
-\rho B_i^{(3)}
\Bigg]\no\\
%%%%%%%%%%%%%%%%%%%%%%%%%%%%%%
T_{ij}^{(3)}\!\!\!\!\!&=&\!\!\!\!\!\frac{1}{6}\Bigg(
2h_{ij}^{(3)}P+6\(h_{ij}^{(2)}\delta P^{(1)}+h_{ij}^{(1)}\delta P^{(2)}\)+6(\delta P^{(1)}+\delta\rho^{(1)})v_i^{(1)}v_j^{(1)}+3(P+\rho)\Big[
v_i^{(1)}B_j^{(2)}+4h_{jk}^{(1)}v^{k(1)}v_i^{(1)}\no\\
&&+v_i^{(1)}v_j^{(2)}+v_j^{(1)}\(B_i^{(2)}+4h_{ik}^{(1)}v^{k(1)}+v_i^{(2)}-8v_i^{(1)}\Psi^{(1)}\)\Big]+3B_j^{(1)}
\Big[
2(B_i^{(1)}+v_i^{(1)})(\delta P^{(1)}+\delta \rho^{(1)})\no\\
&&+(P+\rho)\times\(B_i^{(2)}+4h_{ik}^{(1)}v^{k(1)}+v_i^{(2)}-4B_i^{(1)}\Phi^{(1)}-2v_i^{(1)}(\Phi^{(1)}+2\Psi^{(1)})\)\Big] + 3B_i^{(1)}
\Big[
\no\\
&&2v_j^{(1)}(\delta P^{(1)}+\delta \rho^{(1)})+(P+\rho) \Big(B_j^{(2)}+4h_{jk}^{(1)}v^{k(1)}+v_j^{(2)}-2v_j^{(1)}(\Phi^{(1)}+2\Psi^{(1)})
\Big)
\Big]\no\\
&&+\delta_{ij}\[\delta P^{(3)}-2P\Psi^{(3)}-6(\delta P^{(1)}\Psi^{(2)}+\delta P^{(2)}\Psi^{(1)})\]
\Bigg),
\n
where $v_i\equiv v_i^\perp+\p_iv$ and $B_i$ are the velocity perturbation and vector perturbation in Newton Gauge respectively.

The equation of motion for each $h_{ij}^{(n)}$ takes the same form as Eq.~(\ref{eqh}) and the third-order source term reads
\e
\begin{split}
	\Sij{3} = - \frac{4}{9 \mH^4 \(1+w\)^2} &\Big\{
	\(\mH \p_i\Phi + \p_i\Phi'\) \(\mH \p_j\Phi + \p_j\Phi'\)
	\Big[-2 \(1+\cs\) \p^2\Phi
	+ 6 \mH^2 \(\cs-w\) \Phi 
	\\&+ 3 \mH \(3+2\cs+w\) \Phi'\Big]
	+ 3\mH \(1+w\) \(4\mH \Phi - \Phi'\) \p_i\Phi' \p_j\Phi'
	\\&- 3\mH^3 \(1+w\) \(2\mH \(5+3w\) \Phi - \Phi'\) \p_i\Phi \p_j\Phi
	\Big\},
\end{split}
\q
and the fourth-order source term is given by
\e
\begin{split}
	\Sij{4} = &16 \Phi^3 \p_i\p_j\Phi - \frac{4}{27 \mH^6 \(1+w\)^3}\Big\{ 
	\(\mH \p_i\Phi + \p_i\Phi'\) \(\mH \p_j\Phi + \p_j\Phi'\) \Big[
	4 \(1+\cs\)^2 \(\p^2\Phi\)^2 \\
	&\qquad\quad - \mH^2 \(9+5\cs+9w+9w\cs\) \p_k\phi \p^k\Phi
	+ 4 \cs \(2\mH \p^k\Phi + \p^k\Phi'\) \p_k\Phi'\\
	&\qquad\quad	- 6 \mH \(1+\cs\) \(2 \mH \(3+2\cs+w\) \Phi
	+ \(5+4\cs+w\) \Phi'\) \p^2\Phi \\
	&\qquad\quad + 18 \mH^3 \(7+4c_s^4+7\cs-w\cs+7w+4w^2\) \Phi \Phi'\\
	&\qquad\quad + 36 \mH^4 \(1+c_s^4-2w\cs+2w+2w^2\) \Phi^2
	+ 9 \mH^2 \(1+\cs\) \(5+4\cs+w\) \(\Phi'\)^2
	\Big]\\
	&+ 6 \mH \(1+w\) \p_i\Phi' \p_j\Phi' \Big[
	\(1+\cs\) \( 12 \mH^3 \Phi^2 + \Phi' \p^2\Phi
	- 4 \mH \Phi \p^2\Phi - 3 \mH \(\Phi'\)^2\)
	\\&+ 3 \mH^2 \(-1+3\cs-4w\) \Phi \Phi'\Big]
	 - 3 \mH^3 \(1+w\) \p_i\Phi \p_j\Phi \Big[
	\(1+\cs\) \(2 \Phi' \p^2\Phi - 8 \mH \Phi \p^2\Phi 
	+ 18 \mH^2 \Phi \Phi'\)\\
	&\qquad\quad -3 \mH \(3+2\cs+w\) \(\Phi'\)^2
	+ 6 \mH^3 \(11+4\cs+22w+15w^2\) \Phi^2\Big]
	\Big\}.
\end{split}
\q
Setting $w=\cs=1/3$ during RD, they return to the source terms used in \cite{Yuan:2019udt}.
In our convention, the density parameter for high-order corrections takes the same form as Eq.~(\ref{ogwd}). After defining the Fourier transform of $S_{ij}$ to be
\e
\mS({\bk, \eta}) = -4 e^{ij} S_{ij}(\bk) = \mSi{2} + \mSi{3} +\mSi{4},
\q
then we can write the source terms as
\m\label{sij234}
\mSi{2} &=& 4\int \frac{\rd^3p}{(2\pi)^{3/2}} \be_{\bm{k}}(\bp, \bp)
F^{(2)}(p,|\bk-\bp|,\eta) \Phi_{p} \Phi_{|\bm{k-p}|},\\
\mSi{3} &=& 4\int \frac{\rd^3p \rd^3q}{(2\pi)^3} \be_{\bm{k}}(\bp, \bq)
F^{(3)}(p,q,|\bk-\bp-\bq|,\eta) \Phi_{p} \Phi_{q} \Phi_{|\bm{k-p-q}|},\\
\mSi{4} &=& 4\int \frac{\rd^3p \rd^3q \rd^3l}{(2\pi)^{9/2}} \Big[ 
\be_{\bm{k}}(\bl, \bl) F^{(4)}_1(p,q,l,|\bk-\bp-\bq-\bl|,\eta)
\\
&&\qquad\qquad\qquad+\be_{\bm{k}}(\bp,\bq) F^{(4)}_2(p,q,l,|\bk-\bp-\bq-\bl|,\eta)              
\Big] \Phi_{p} \Phi_{q} \Phi_{l} \Phi_{|\bm{k-p-q-l}|},\qquad
\n
where we have defined $\be_{\bm{k}}(\bp, \bq) \equiv e^{ij}(\vec{k})p_iq_j$. The contribution for the cross mode is omitted here since the plus mode, and cross mode have the same energy density and we can sum over the polarization mode at the final step. In Eq.~(\ref{sij234}), the time evolution of $\Phi$ is absorbed in the transfer function, namely
\e
\begin{split}
F^{(2)}(\bq_1, \bq_2,\eta) &= \frac{1}{\mH^2} \(
3\mH^2 T_{q_1} T_{q_2} + \mH (T_{q_1} T'_{q_2} + T'_{q_1} T_{q_2})
+ T'_{q_1} T'_{q_2}\),\\
F^{(3)}(\bq_1, \bq_2, \bq_3,\eta) &= \frac{1}{3 \mH^4} \Big\{
2 \mH^2 T_{q_1} T_{q_2} \[\(18 \mH^2 - q_3^2\) T_{q_3} - 6 \mH T_{q_3}'\]
- 2 T_{q_1}' T_{q_2}' \[\(6 \mH^2 + q_3^2\) T_{q_3} + 3 \mH T_{q_3}'\]\\
&\quad - \mH \(T_{q_1} T_{q_2}' + T_{q_1}' T_{q_2}\) 
\(2 q_3^2 T_{q_3} + 9 \mH T_{q_3}'\)\Big\},\\
F^{(4)}_1(\bq_1, \bq_2, \bq_3, \bq_4,\eta) &= 16\, T_{q_1} T_{q_2} T_{q_3} T_{q_4},\\
F^{(4)}_2(\bq_1, \bq_2, \bq_3, \bq_4,\eta) &= \frac{1}{36 \mH^6} \Big\{
\mH^2 T_{q_1} T_{q_2} \Big[
- 108 \mH^2 T'_{q_3} T'_{q_4} 
+ 48 \mH^2 \(24\mH^2 + q_3^2 + q_4^2\) T_{q_3} T_{q_4}\\ &\qquad\qquad\qquad + 12 \mH \(9\mH^2 - q_3^2\) T_{q_3} T'_{q_4} 
+ 12 \mH \(9\mH^2 - q_4^2\) T'_{q_3} T_{q_4} 
\Big]\\
&\qquad\quad + T'_{q_1} T'_{q_2} \Big[
72 \mH^2 T'_{q_3} T'_{q_4} 
- 48 \mH^2 \(6\mH^2 + q_3^2 + q_4^2\) T_{q_3} T_{q_4}\\
&\qquad\qquad\qquad + 12 \mH \(3\mH^2 + q_3^2\) T_{q_3} T'_{q_4} 
+ 12 \mH \(3\mH^2 + q_4^2\) T'_{q_3} T_{q_4} 
\Big]\\
&\qquad\quad - \(\mH T_{q_1} + T'_{q_1}\) \(\mH T_{q_2} + T'_{q_2}\) \Big[
12 \mH \(21\mH^2 + 5 q_3^2\) T_{q_3} T'_{q_4}\\
&\qquad\qquad\qquad + 12 \mH \(21\mH^2 + 5 q_4^2\) T'_{q_3} T_{q_4}
+ 180\mH^2 T'_{q_3} T'_{q_4}\\
&\qquad\qquad\qquad + 8 \(18\mH^4 + 9\mH^2 q_3^2 + 9\mH^2 q_4^2 
+ 2 q_3^2 q_4^2\) T_{q_3} T_{q_4}\\
&\qquad\qquad\qquad +3 \bq_3 \cdot \bq_4 \(11\mH^2 T_{q_3} T_{q_4}
- \mH T_{q_3} T'_{q_4} - \mH T'_{q_3} T_{q_4} - T'_{q_3} T'_{q_4}\)
\Big]\Big\},
\end{split}
\q
where we set $w=\cs=1/3$.
The notation $T_{k}$ is short for the time evolution function, $T_\Phi(k\eta)$. Applying the Green's function method, the solution to $h_{ij}^{(n)}$ takes the same form as Eq.~(\ref{solh}). The next-order corrections to $\Omega_{\mathrm{GW}}$ comes from two parts. The first part is the coupling of $S_{ij}^{(3)}$ itself,
\e\label{hh}
	\begin{split}
		\Omega_{\mathrm{GW}}^{\uppercase\expandafter{\romannumeral1}}(k,\eta) =& \frac{k^3}{384\pi^2}\({k\over\mH}\)^2\int \rd \te_1 \frac{a(\te_1)}{a(\eta)} g_k(\eta; \te_1)
		\int \te_2 \frac{a(\te_2)}{a(\eta)} g_k(\eta, \te_2)
		\int \rd^3p \rd^3q \frac{P_\Phi(\mpq)}{\mpq^3} 
		\frac{P_\Phi(q)}{q^3} \frac{P_\Phi(\mkp)}{\mkp^3}\\
		&\quad \times \Big[2 \bm{e}_{\bm{k}}(\bp-\bq, \bq)^2 F^{(3)}(q,\mpq,\mkp, \te_1)
		F^{(3)}(q,\mpq,\mkp,\te_2)\\
		&\qquad+ 4 \bm{e}_{\bm{k}}(\bp,\bp-\bq) 
			\bm{e}_{\bm{k}}(\bp,\bq) 
		F^{(3)}(\mkp,\mpq,q,\te_1)
		F^{(3)}(q,\mkp,\mpq,\te_2)\Big],
	\end{split}
\q
while the second part is the coupling of $S_{ij}^{(2)}$ and $S_{ij}^{(4)}$
\e\label{hh2}
	\begin{split}
		\Omega_{\mathrm{GW}}^{\uppercase\expandafter{\romannumeral2}}(k,\eta) =& \frac{k^3}{384\pi^2}\({k\over\mH}\)^2\int \rd \te_1 
		\frac{a(\te_1)}{a(\eta)} g_k(\eta; \te_1)
		\int \te_2 \frac{a(\te_2)}{a(\eta)} g_k(\eta, \te_2)
		\int \rd^3p \rd^3q \frac{P_\Phi(p)}{p^3} 
		\frac{P_\Phi(q)}{q^3} \frac{P_\Phi(\mkp)}{\mkp^3}\\
		& \times 
		\be_{\bm{k}}(\bp,\bp) F^{(2)}(\bp,\bk-\bp,\te_1) \Big\{
		6\[\be_{\bm{k}}(\bp,\bp) + \be_{\bm{k}}(\bq,\bq)\] F_1^{(4)}(\bp,\bk-\bp,\bq,\bq,\te_2)\\ 
		&-2 \be_{\bm{k}}(\bp,\bp) F_2^{(4)}(\bp,\bk-\bp,\bq,-\bq,\te_2) -8\be_{\bm{k}}(\bp,\bq) F_2^{(4)}(\bp,\bq,\bq,\bk-\bp,\te_2)\\
		&-2 \be_{\bm{k}}(\bq,\bq) F_2^{(4)}(\bq,\bq,\bp,\bk-\bp,\te_2)
		\Big\}.
	\end{split}
\q

\begin{figure}[htbp!]
	\centering
	\includegraphics[width = 0.6\textwidth]{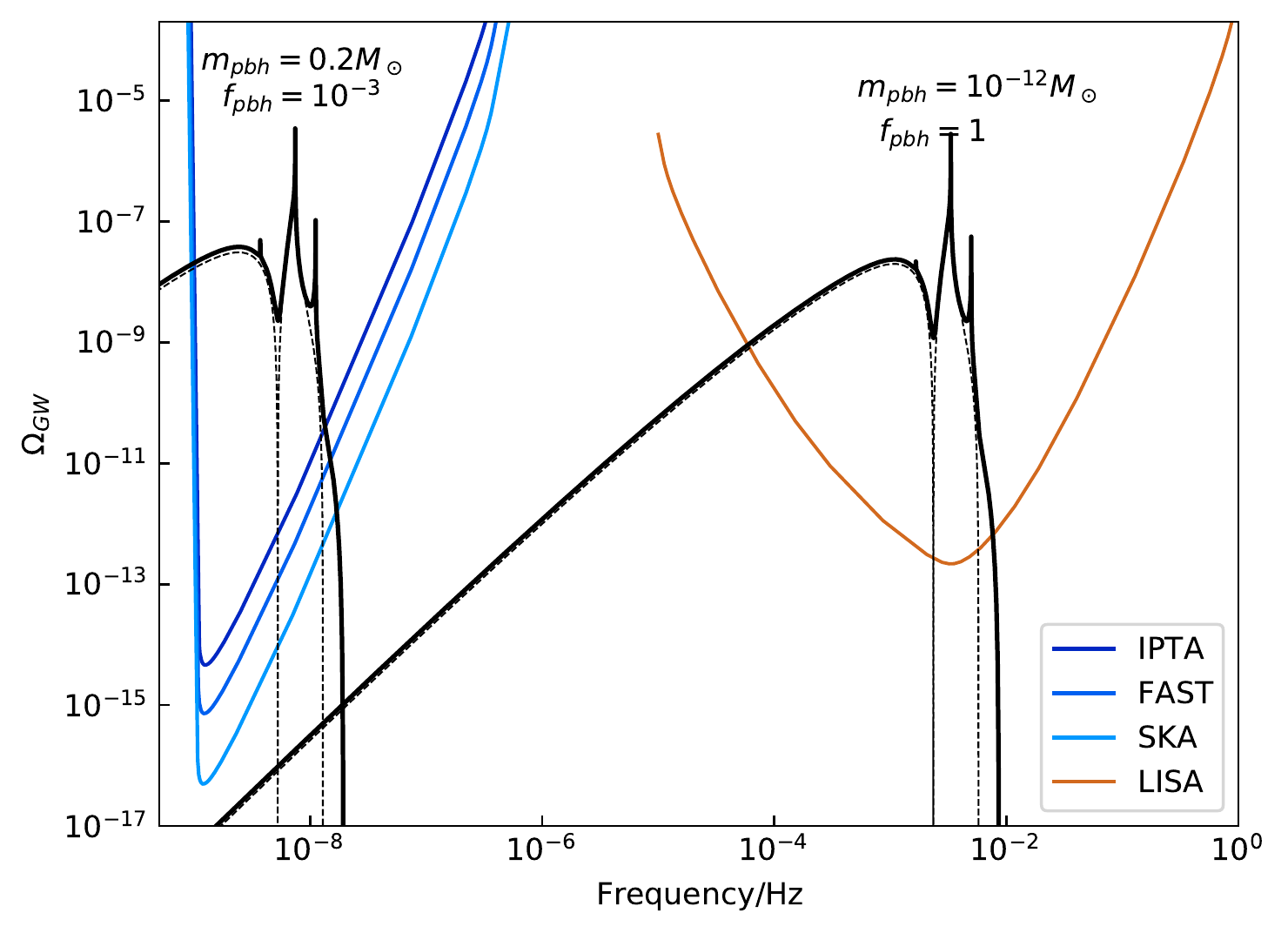}
	\caption{\label{3rdogw} 
		Taken from Fig.~1 in \cite{Yuan:2019udt}. 
		The energy density parameter of SIGWs generated by a $\delta$-spectrum up to third-order. The sensitivity curves for LISA, IPTA, FAST and SKA are shown.
		The dashed lines denote the second-order SIGWs, and the solid black lines include the third-order correction.
	}
\end{figure}

The next-order corrections of $\Omega_{\mathrm{GW}}$ is the sum of Eq.~(\ref{hh}) and Eq.~(\ref{hh2}).
For a $\delta$-spectrum, these two expressions can be further simplified after integrate over the $\delta$ function (see \cite{Yuan:2019udt} for results). After considering the higher-order corrections, the deep valley at $k=\sqrt{2/3}k_*$ generated at second-order will be smoothed to a finite value. Moreover, the cut-off frequency will be extended from $2k_*$ to $3k_*$. See Fig.~\ref{3rdogw}.

The above calculation only focus on the GWs induced by the linear-order scalar perturbations. However, the higher-order scalar, vector and tensor perturbations will contribute to the higher-order SIGWs. Recently, the SIGWs induced by the second-order perturbations were analytically studied in \cite{Zhou:2021vcw}. But the complete calculation of the next order correction to SIGWs has not been explored yet.

\subsection{Scalar Induced Gravitational Waves within Non-Gaussianities}
GWs generated by non-Gaussian scalar perturbations were first estimated by Nakama \cite{Nakama:2016gzw} where they found that the amplitude of SIGWs could be suppressed by non-Gaussianities to several orders of magnitude. After \cite{Espinosa:2018eve,Kohri:2018awv} proposed the semi-analytical method to calculate SIGWs, \cite{Cai:2018dig,Unal:2018yaa} calculate the non-Gaussian effects on SIGWs by considering a local-type non-Gaussianities up to second-order (or $\Fnl$-order). They both argued that SIGWs within non-Gaussianities have observable signatures which can be used to probe primordial non-Gaussianities.
However, in a more recent study \cite{Yuan:2020iwf}, Yuan {\it et al.} revisited the non-Gaussian effects and extended the calculation to third-order  (or $\Gnl$-order non-Gaussianities). They argue that all the non-Gaussian effects are degenerate with the power spectrum. Hence it is impossible to read any information about non-Gaussianities only through the signal of SIGWs.

To see this more clear, let us go back to the first line of Eq.~(\ref{ogw(k)}), where one needs to compute the two-point correlator of the source terms. This will lead to the four-point correlator of $\Phi$, namely $\<\Phi_{\bm{p}}\Phi_{\bm{k-p}}\Phi_{\bm{q}}\Phi_{\bm{k'-q}}\>$. If $\Phi$ obeys Gaussian distribution, then the four-point correlator can be simplified to three two-point correlators through Wick's theorem. However, the general formula for the four-point correlator is
\m\label{4-point}
\<\Phi_{\bm{p}}\Phi_{\bm{k-p}}\Phi_{\bm{q}}\Phi_{\bm{k'-q}}\>
&=&\<\Phi_{\bm{p}}\Phi_{\bm{k-p}}\Phi_{\bm{q}}\Phi_{\bm{k'-q}}\>_c
+\<\Phi_{\bm{p}}\Phi_{\bm{k-p}}\>\<\Phi_{\bm{q}}\Phi_{\bm{k-q}}\>
+\<\Phi_{\bm{p}}\Phi_{\bm{q}}\>\<\Phi_{\bm{k-p}}\Phi_{\bm{k'-q}}\>\no\\
&+&\<\Phi_{\bm{p}}\Phi_{\bm{k'-q}}\>\<\Phi_{\bm{k-p}}\Phi_{\bm{q}}\>
\n
where we assume that $\<\Phi\>=0$ and $\<\Phi_{\bm{p}}\Phi_{\bm{k-p}}\Phi_{\bm{q}}\Phi_{\bm{k'-q}}\>_c$ is the connected four-point correlation function (4PCF) (or called the fourth cumulant in statistics) which is related to the trispectrum, $\mathcal{T}_\Phi$, by 
\e
\<\Phi_{\bm{p}}\Phi_{\bm{k-p}}\Phi_{\bm{q}}\Phi_{\bm{k'-q}}\>_c=(2\pi)^3\delta(\bm{k}+\bm{k'})\mathcal{T}_\Phi(\bm{p},\bm{q},\bm{k},\bm{k'}).
\q

The connected 4PCF will vanish if $\Phi$ is a Gaussian variable. Otherwise it will contribute to be SIGWs \cite{Unal:2018yaa}. In \cite{Unal:2018yaa,Cai:2018dig}, the authors consider a local-type non-Gaussianities up to $\Fnl$-order and calculate the SIGWs in the absence of the connected 4PCF. After that, the $\Gnl$-order is studied by \cite{Yuan:2020iwf} still in the absence of the connected 4PCF. 
Recently, \cite{Atal:2021jyo} considered the leading order of the connected 4PCF and \cite{Adshead:2021hnm} included the complete contribution of the connected 4PCF.
Although non-Gausiannities can alter the waveform of the SIGW signal, one can not read any information about non-Gaussianities only through the signal of $\Omega_{\mathrm{GW}}(f)$. This is because the non-Gaussian effects are absorbed into the total 4PCF and there will be a degeneracy between the non-Gaussian effects and the total 4PCF. Therefore, one can not tell from the total 4PCF whether there is a evidence for non-Gaussianities (or for the connected 4PCF).

\subsection{Gauge issue of Scalar Induced Gravitational Waves}
Although the tensor mode are gauge invariant at first-order, they fail to remain gauge invariant at second-order (see, e.g., \cite{Noh:2003yg}). Therefore, a natural question arises: Is the energy density of SIGWs gauge dependent or not?

For a long time, Newton gauge has been a commonly used gauge to compute SIGWs since the degrees of freedom are fixed completely in this gauge, and there is no residual gauge freedom. In Newton gauge, the scalar mode $\phi$ is the just Bardeen potential $\Phi$ and all the degrees of freedom are fixed, which make Newton gauge the most mathematically convenient gauge for evaluating SIGWs. SIGWs in other gauges were first calculated numerically by Hwang {\it et al.} \cite{Hwang:2017oxa} where they investigated $\Omega_{\mathrm{GW}}$ in uniform expansion gauge, comoving gauge and uniform curvature gauge and they found that $\Omega_{\mathrm{GW}}$ is gauge dependent. After that, the semi-analytical method was developed, and Gong calculated the SIGWs in comoving gauge during MD \cite{Gong:2019mui}. Gong's result showed that $h_{ij}$ increases as $\eta^2$ in comoving gauge. Another study by Tomikawa and Kobayashi appeared at the same time \cite{Tomikawa:2019tvi}, where they investigate the SIGWs in comoving gauge and uniform curvature gauge. They found that the results in comoving gauge increase with time for $w\ge0$ while the result in uniform curvature gauge is identical with that in Newton gauge for $w>0$. Moreover, they found that the results are all different in Newton gauge, comoving gauge, and uniform curvature gauge if $w=0$.

Due to the gauge dependence of $\Omega_{\mathrm{GW}}$, it is natural to ask which gauge is relevant in interpreting the observations. This question is discussed by De Luca {\it et al.} \cite{DeLuca:2019ufz}. They calculated the SIGWs in synchronous gauge since they argued that the sensitivity curves of LISA are given in that gauge. By neglecting the $E$ mode but keeping its derivatives in the source term , De Luca {\it et al.} found that the SIGWs in synchronous gauge are identical to that of Newton gauge. Their results are later confirmed by \cite{Yuan:2019fwv,Inomata:2019yww}.
However, the calculation of SIGWs in synchronous gauge is a a tough nut since there is residual gauge freedom in $E$ mode. The complete calculation in synchronous gauge and which gauge is relevant for observations are still open questions.

In synchronous gauge, $\delta g_{00}=\delta g_{0i}=0$, corresponding to $\tilde{\phi}=\tilde{B}=0$ in Eq.~(\ref{1st-GT}). This leads to \cite{Lu:2020diy}
\m
T^{S}(\eta)&=&-{1\over a}\(\int_0^\eta a(\tilde{\eta})\phi(\tilde{\eta})\mathrm{d}\tilde{\eta}-\mathcal{C}_1(\bm{x})\)\\
L^{S}(\eta)&=&\int_0^\eta[T^{S}(\tilde{\eta})-B(\tilde{\eta})]\mathrm{d}\tilde{\eta}+\mathcal{C}_2(\bm{x}),
\n
where $\mathcal{C}_1$ and $\mathcal{C}_2$ are two arbitrary spatial functions. The presence of $\mathcal{C}_1$ and $\mathcal{C}_2$ come from the residual gauge freedom in synchronous gauge. To determine the time slicing and the spatial coordinate on the hypersurface, one has to fix $\mathcal{C}_1$ and $\mathcal{C}_2$. To see the impacts of residual gauge freedom more clearly, we derive the equation of motion for the scalar modes during RD, namely
\m
2\mH E'+E''+\psi&=&0,\\
6\psi''+2\mH\(9\psi'-4\p^2E\)-3\p^2 E''-5\p^2\psi&=&0.
\n
After some algebra, one can get the equation of motion for $E$, (see also \cite{Lu:2020diy})
\e\label{eqh-E}
x^3 T_E''''(x)+5x^2 T_E'''(x)+\(2+{x^2\over3}\)xT_E''(x)-\(2-{x^2\over 3}\)T_E'(x)=0,
\q
where the transfer function for $E$ and $\psi$ are defined as
\m
k^2E(\vec{k})&\equiv&
\Phi_kT_E(k\eta)\\
\psi(\vec{k})&\equiv&\Phi_k
T_\psi(k\eta).
\n
In our convention, the boundary condition for the transfer function should be $T_\psi(x\to0)=3/2$ in order to match the relation between curvature perturbation and the Bardeen potential.
The general solution for Eq.~(\ref{eqh-E}) is 
\e\label{TE}
T_E(x)=\mathcal{C}_3+\mathcal{C}_4\(\Ci(z)-{\sin(z)\over z}\)+\mathcal{C}_5\ln(z)+\mathcal{C}_6\(\Si(z)+{\cos(z)\over z}\),
\q
where $z\equiv x/\sqrt{3}$ and $\mathcal{C}_i,(i=3,4,5,6)$ are integrate constants.
One can check that under the residual gauge transformation,
\m\label{residual}
T&=&-{\mathcal{C}_7\over x}\\
L&=&-\mathcal{C}_7 \ln x +\mathcal{C}_8,\label{residual1}
\n
the gauge condition $\tilde\phi=\tilde B=0$ still satisfy. Combining Eq.~(\ref{TE}) and Eq.~(\ref{residual1}), we see that $\mathcal{C}_8$ can be absorbed by $\mathcal{C}_3$ while $\mathcal{C}_7$ corresponds to $\mathcal{C}_5$. In other words, the constant $\mathcal{C}_3$ and the logarithm term are two independent pure gauge modes. One can arbitrarily choose $\mathcal{C}_3$ and $\mathcal{C}_5$ to work in a specific hypersurface. After selecting the hypersurface, the solution can be well determined after performing the only physical condition, $T_\psi(x\to0)=3/2$.
Here comes a big problem in synchronous gauge. The solution to $E$ mode will definitely diverge either at the initial time or at a late time.
For instance, let's assume that $T_E(x)$ have a finite value at $x=0$. Notice that the Taylor expansion of cosine integral function at $x=0$ is $\Ci(x)=\gamma+\ln(x)+\mathcal{O}(x^2)$, where $\gamma$ is the Euler gamma constant. Therefore, one must choose $\mathcal{C}_5=-\mathcal{C}_4$ so that the logarithm term can be cancelled by the $\Ci(x)$ term. By applying the boundary condition $T_\psi(0)=3/2$, one gets $\mathcal{C}_6=0$ and $\mathcal{C}_4=-\mathcal{C}_5=9$ and the finite value at $x=0$ is given by $T_E(0)=\mathcal{C}_3+9\gamma$. However, under this situation, we see that the $T_E(x)$ at a late time will diverge according to the logarithm term in Eq.~(\ref{TE}). The consequence of the divergence is that the $E$ mode will induce the GWs continuously during RD, resulting in the divergence of $\Omega_{\mathrm{GW}}$ for SIGWs. On the other hand, if we let the gauge mode $\mathcal{C}_5$ vanish, the presence of the $\Ci(x)$ term will result in the divergence at the initial time. A physical interpretation is that if the observer requires a converged initial condition, the geodesics of the observer using as a reference may collapse or cross each other. So one has to change the residual gauge freedom to adjust the reference system. This makes the calculation of SIGWs in synchronous gauge very complicated.

In \cite{DeLuca:2019ufz,Yuan:2019fwv,Inomata:2019yww}, the authors calculated the SIGWs in the absence of $E$ mode but considering its derivatives where the gauge freedom are completely fixed. In this situation, the energy density of SIGWs agrees with that in Newton gauge. Moreover, it was shown in \cite{Lu:2020diy} that if throwing away the pure gauge modes, the result returns to that in Newton gauge. These studies indicate that the large divergence of SIGWs in synchronous gauge comes from the residual gauge freedom and the synchronous gauge seems to be ill-defined for calculating the SIGWs.
Readers interested in the calculation of different gauge can refer to \cite{Lu:2020diy,Ali:2020sfw,Inomata:2020cck}.

Since the standard procedure to calculate the energy density of $h_{ij}$ brings divergence, one way to tackle the problem is to find a new gauge invariant quantity for the GWs. One can construct infinite gauge invariant variables for scalar and tensor perturbations. However, the point is to find a quantity that can interpret the physical world and the measurements.
This idea has been tried by Zhang, Wang and Zhu in \cite{Chang:2020tji,Chang:2020iji}. By applying the technique of Lie derivative, they constructed gauge invariant second-order GWs in synchronous gauge, which is related to the measurement, and their results coincide with that computed in Newton gauge. They also used this idea to evaluate the energy density of SIGWs in uniform density gauge, which is supposed to diverge as $\eta^6$. They found the new gauge invariant second-order GWs converged and were identical with the one in Newton gauge \cite{Chang:2020mky}.
In a more recent study \cite{Domenech:2020xin}, it is shown that the current $\Omega_{\mathrm{GW}}$ is well defined in most of gauges if one takes the sub-horizon limit, namely $\eta\to\infty$. The authors found that $\Omega_{\mathrm{GW}}$ will be the same as that in Newton gauge if the trace part of the metric perturbation at sub-horizon limit is the order of or smaller than the scalar perturbation $\Phi$ in Newton gauge (see Eq.~(3.2) or Eq.~(3.5) in \cite{Domenech:2020xin}). They also pointed out that, for particular choices of the residual gauge freedom, synchronous gauge can give the same result in Newton gauge.

Although the synchronous gauge is relevant for observations, Newton gauge is still the most popular gauge since it not only gives the same result as synchronous gauge but also is the simplest gauge to perform the calculation.

\section{Searching for Primordial Black Hole Dark Matter Using Scalar Induced Gravitational Waves}

SIGWs provide the most efficient way so far to search for PBH DM. This is due to the fact that PBHs might from the peak of scalar perturbations, which is extremely sensitive to the amplitude of the power spectrum, $A$.
Let's consider a monochromatic power spectrum for example, described by Eq.~(\ref{delta}). The variance of the perturbations generated by this spectrum is $\sigma=\int P_\zeta(k)\mathrm{d}\ln k=A$. Using the Press-Schechter formalism, Eq.~(\ref{PSbeta}), the mass function can be approximated as
\e
\beta=\half\mathrm{erfc}\({\nu_c\over\sqrt{2}}\)=\sqrt{1\over 2\pi}{\mathrm{e}^{-\nu_c^2/2}\over\nu_c}\propto \sqrt{A}~\mathrm{e}^{-{\Delta_c^2\over2A}},
\q
which is exponentially dependent on $A$.
On the other hand, the inevitably generated SIGWs satisfy $h_{ij}\propto A$.
Therefore, a small change in $A$ would change $\fpbh$ by several orders of magnitude.

\begin{figure}[htbp!]
	\centering
	\includegraphics[width = 0.7\textwidth]{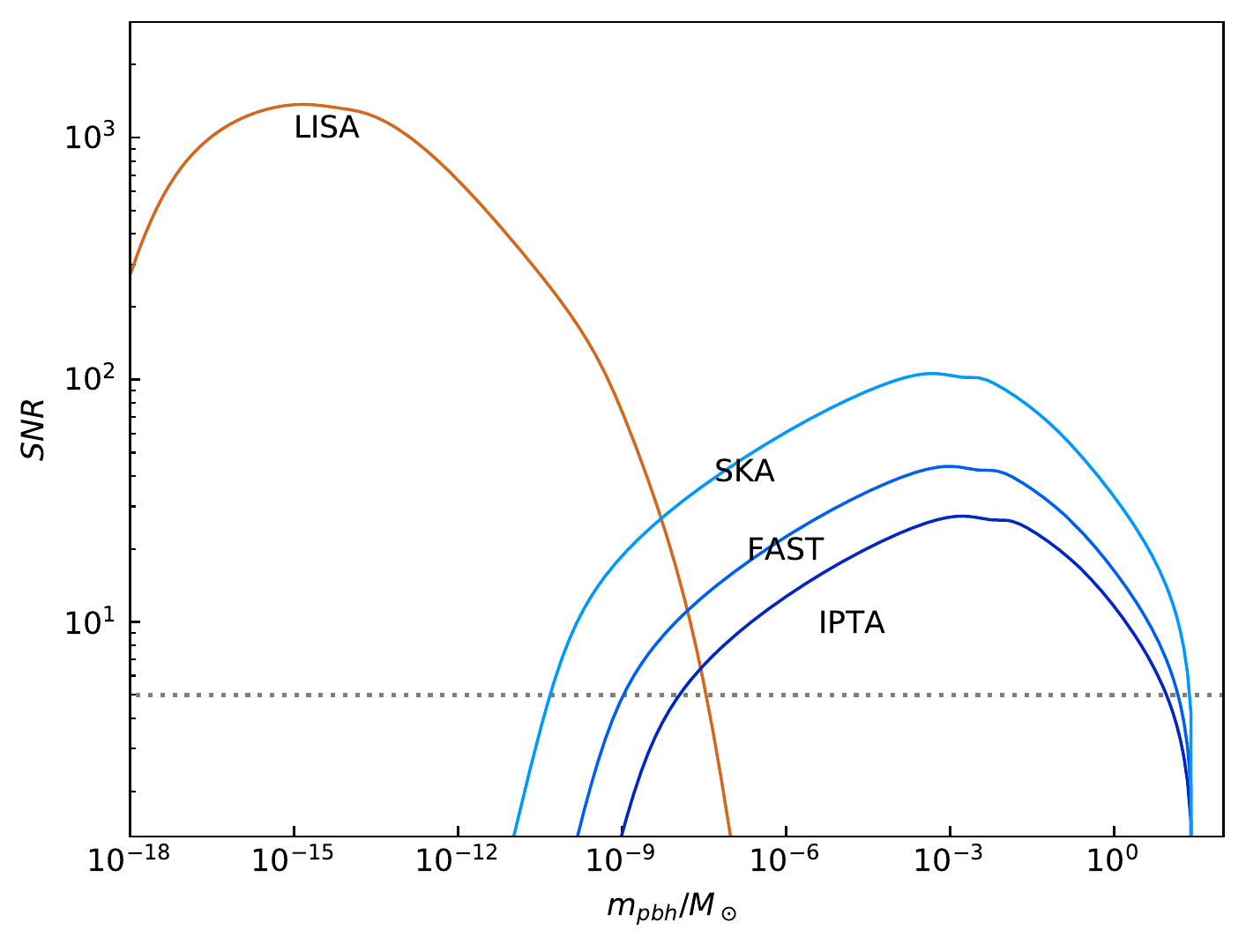}
	\caption{\label{SNRLISA} The expected SNR by LISA/IPTA/FAST/IPTA in detecting SIGWs generated by monochromatic PBHs. The dotted line corresponds to $\SNR=5$. We assume the observation time to be $T=4$yr for LISA and $T=30$yr for PTAs. The value of $\Delta t$, $N$ and $\sigma$ for current PTAs can be found in \cite{Kuroda:2015owv} (see Table. 5).
	}
\end{figure}

To further quantify the power of SIGWs, we shall estimate the expected signal-to-noise ratio (SNR) by the GW detector, which is evaluated as \cite{Allen:1997ad,Thrane:2013oya}, 
\e\label{SNR}
\rho^{2}\!=T \! \int \! \mathrm{d} f \frac{\Gamma(f)^{2} S_{h}(f)^{2}}{\left[{1\over 25}+\Gamma(f)^{2}\right] S_{h}(f)^{2}+P_{n}(f)^{2}+{2\over 5} S_{h}(f) P_{n}(f)},
\q
The strain power spectral density is given by $S_h(f) = 3H_0^2\Omega_{\mathrm{GW},0}/(2\pi^2 f^3)$. $\Gamma(f)$ is the overlap function and $P_n(f)$ is the noise power spectral density. For LISA, $\Gamma(f)=R(f)$ where $R(f)$ is the signal transfer function of LISA and its expression should be computed numerically \cite{Larson:1999we} but it can be well fit by \cite{Cornish:2018dyw}
\e
R(f)={3\over 10(1+0.6(f/f_{\star})^2)},
\q
where $f_\star =c/(2\pi L)= 19.09~\mathrm{mHz}$ is the transfer frequency and $L=2.5~\mathrm{Gm}$ for the current LISA design.
On the other hand, $P_n(f)$ can be approximated by \cite{Cornish:2018dyw}
\e
P_n(f)={P_{\mathrm{oms}}\over L^2}+2(1+\cos^2(f/f_{\star})){P_{\mathrm{acc}}\over{(2\pi f)^4L^2}}.
\q
The optical metrology noise spectrum and the acceleration noise is parameterized by \cite{Cornish:2018dyw}
\m
P_{\mathrm{oms}}(f)&=&(1.5\times10^{-11}\mathrm{m})^2
\(1+\({2 \mathrm{mHz} \over f}\)^4\) \mathrm{Hz}^{-1}
\\
P_{\mathrm{acc}}(f)&=&(3\times 10^{-15}~\mathrm{m~s^{-2}})^2
\(1+\({0.4 \mathrm{mHz} \over f}\)^2\)
\(1+\({f \over 8\mathrm{mHz} }\)^4\)
\mathrm{Hz}^{-1}.
\n
For pulsar timing arrays (PTA), the expressions are $\Gamma(f)=R(f)=1/(12\pi^2f^2)$ and $P_n(f) =2\Delta t\sigma^2$ where $\Delta t$ is the observation time and $\sigma$ is the root-mean-square timing noise.
For PTAs, each of the millisecond pulsars can be regarded as a single GW detector, and one has to average the spatial contribution such that
\cite{Allen:1997ad,Siemens:2013zla} 
\e 
\rho^2 = 2T \sum_{I, J}^{M} \zeta_{IJ}^2\[
\int  \!\! \mathrm{d}f
\frac{R(f)^{2} S_{h}(f)^{2}}{\left[{1\over 25}+R(f)^{2}\right] S_{h}(f)^{2}+P_{n}(f)^{2}+{2\over 5} S_{h}(f) P_{n}(f)}
\],
\q
where $\zeta_{IJ}$ is the normalized Hellings and Downs coefficient for pulsars $I$ and $J$. Assuming the pulsars are distributed homogeneously on the sky, then $\zeta_{IJ}$ takes the form
\e
\zeta_{IJ}={N(N-1)}\frac{4.74}{20\times19},
\q
where $N$ is the number of pulsars.
Combing the above equations, we can estimate the expected SNR from monochromatic PBHs and scan the mass of PBHs. The result is shown in Fig.~\ref{SNRLISA}. PBHs lighter than $\lesssim 10^{-18}\Msun$ have already evaporated through Hawking radiation by today and is not shown in the figure.

\begin{figure}[htbp!]
	\centering
	\includegraphics[width = 0.6\textwidth]{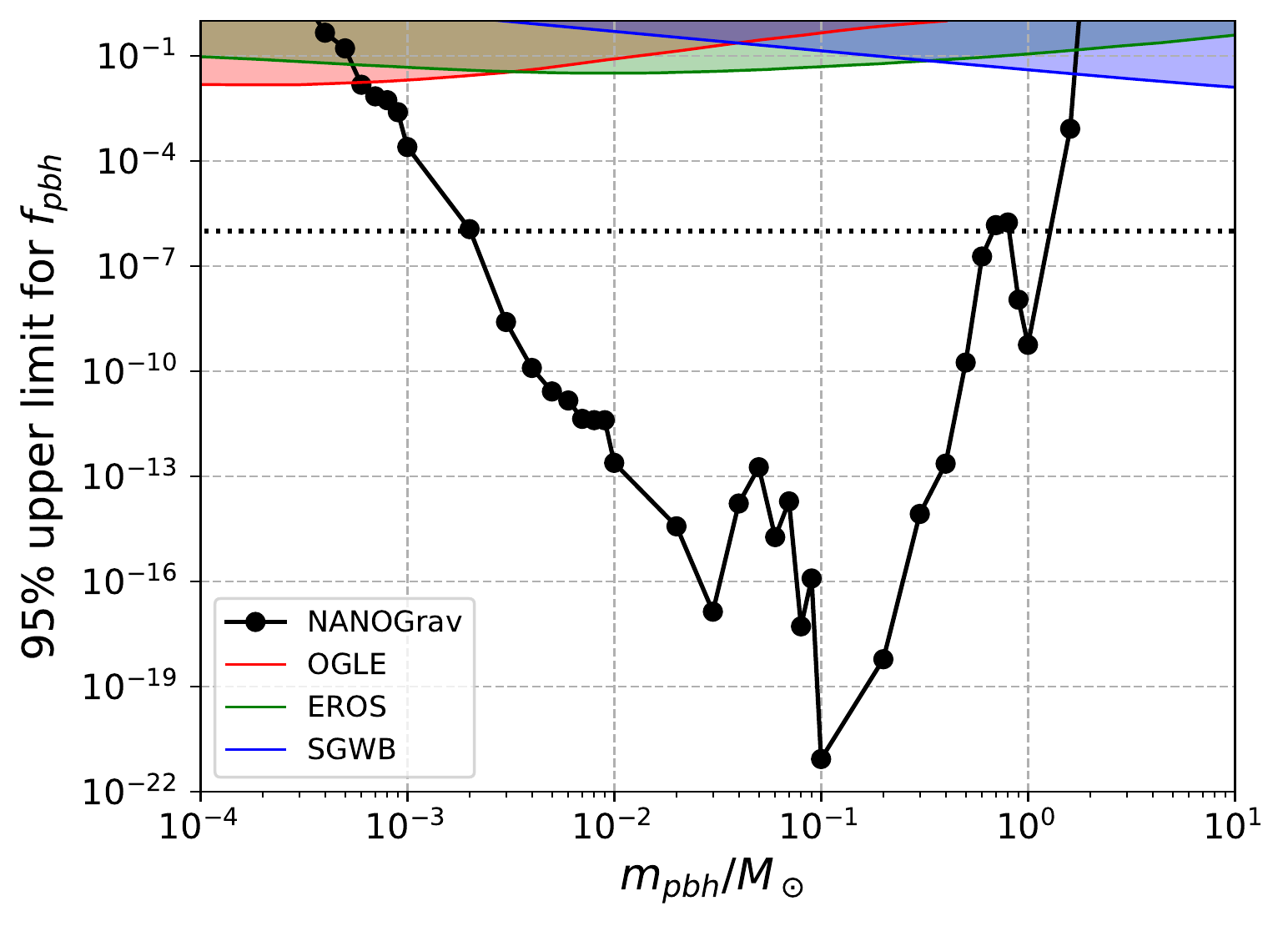}
	\caption{\label{fpbh_upper} 
		Taken from Fig.2 in \cite{Chen:2019xse}. The constraints on $\fpbh$ through the null detection of SIGWs by NANOGrav 11-yr observations. The dotted line corresponds to $10^{-6}$. The colored region are the results excluded by OGLE microlensing \cite{Niikura:2019kqi}, EROS microlensing \cite{Tisserand:2006zx} and the SGWB from PBH binaries \cite{Chen:2019irf}.
	}
\end{figure}

If the PTAs and LISA fail to detect the SIGWs from PBHs, an upper limit can be placed for $\fpbh$. In \cite{Wang:2019kaf}, Wang {\it et al.} found the constraints on $\fpbh$ can reach $10^{-13}$ for $\mpbh\in[10^{-8},1]\Msun$.
Chen {\it et al.} \cite{Chen:2019xse} searched the SIGW signals in the NANOGrav 11-yr data set for monochromatic PBHs. They scan the mass of PBHs and dit not find the SIGW signal. Therefore they placed an upper limit on $\fpbh$, see Fig.~\ref{fpbh_upper}. As shown in the figure, the constraints from SIGW are several orders of magnitude better than the other constraints. Recently, NANOGrav has reported strong evidence for a common-spectrum process modelled as power-law in the 12.5-year dataset. Various models have been put forward to explain the signal assumed to be gravitational waves. SIGWs generated by a broad and flat power spectrum seems to give one possible explanation \cite{DeLuca:2020agl,Sugiyama:2020roc} among the various cosmological and astrophysical models. See also \cite{Vaskonen:2020lbd,Kohri:2020qqd,Domenech:2020ers,Inomata:2020xad}. However, further analysis indicates that the signal is preferred to be scalar transverse modes in the general metric theory rather than the gravitational waves predicted in general relativity \cite{Chen:2021wdo}. 

Another study was made by Kapadia {\it et al.} \cite{Kapadia:2020pir} where they explored the SIGW signal in the LIGO data. Kapadia {\it et al.} reported a null detection of SIGW signal and thus placing a severe constraint on $\fpbh$ to be less than a few parts in million. Their work provides another independent test in addition to the $\gamma$-ray burst \cite{Carr:2009jm} in this mass window.

\begin{figure}[htbp!]
	\centering
	\includegraphics[width = 0.6\textwidth]{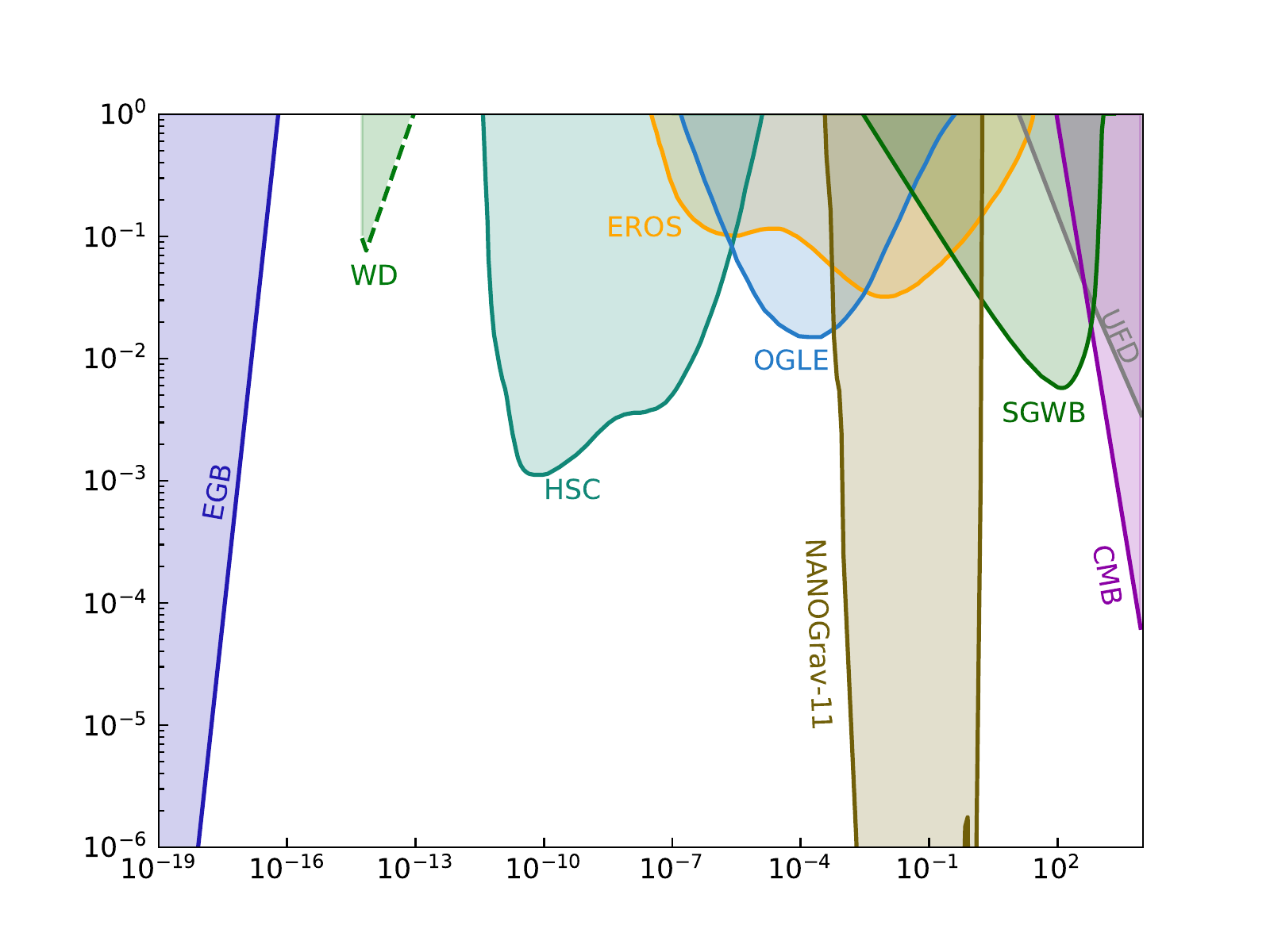}
	\caption{\label{fpbh-c} An overview on the current observational constraints on $\fpbh$. Constraints are taken from EGB \cite{Carr:2009jm}, WDs \cite{Graham:2015apa}, HSC \cite{Niikura:2017zjd}, EROS \cite{Tisserand:2006zx}, OGLE \cite{Niikura:2019kqi}, SGWB from binary PBHs \cite{Wang:2016ana,Chen:2019irf}, UFD \cite{Brandt:2016aco} and SIGWs using NANOGrav 11-yr data set \cite{Chen:2019xse}
	}
\end{figure}

Up to now there are various independent cosmological observations constraining the fraction of PBHs in Fig.~\ref{fpbh-c}. For  $\mpbh\lesssim 10^{-18}\Msun$, PBHs are severely constrained through the null detections of extra-galactic Gamma-ray background (EGB) from PBH evaporation \cite{Carr:2009jm}. On the other hand, PBHs heavier than $\sim 10^{3}\Msun$ are tightly constrained by the CMB observations through the accreting PBHs \cite{Ali-Haimoud:2016mbv,Blum:2016cjs,Horowitz:2016lib,Chen:2016pud}. In the mass range $\mpbh\in[10^{-18},10^3]\Msun$, $\fpbh$ are constrained to no more than a few in thousand by numerous astrophysical observations. For example, 
microlensing events such as Subaru/HSC \cite{Niikura:2017zjd}, Kepler \cite{Griest:2013esa}, OGLE \cite{Niikura:2019kqi} and EROS/MACHO \cite{Tisserand:2006zx} and the dynamical heating of ultra-faint dwarf galaxies \cite{Brandt:2016aco}.
On the other hand, GWs physics also placed broad constraints on $\fpbh$, such as the merger events from binary PBHs in the subsolar mass range \cite{Abbott:2018oah,Magee:2018opb,Chen:2019irf,Authors:2019qbw} and the null detections of stochastic GW background (SGWB) from binary PBHs with LIGO \cite{Wang:2016ana,Chen:2019irf}.
Although the existence of white dwarfs (WDs) in our local galaxy have also placed constraints on $\fpbh$ \cite{Graham:2015apa}. However, this constraint has been challenged by a recent paper \cite{Montero-Camacho:2019jte}, and we label the constraint from WDs with the dashed line in Fig.~\ref{fpbh-c}. Despite the above constraints, PBHs in a substantial window $\sim[10^{-16},10^{-14}] \cup [10^{-13},10^{-12}] M_\odot$ are still allowed to account for all of the DM.

Although SIGWs place the currently severest constraint on $\fpbh$, it is based on the standard PBH formation model in which PBHs are formed from the critical collapse of overdensity. However, other mechanism of the PBH formation has been proposed (see e.g., \cite{Cotner:2019ykd}) and the constraints from SIGWs are not valid for those PBHs.

\section{Summary and Outlook}
PBHs can represent the DM in our Universe and explain the merger events detected by LIGO/VIRGO if $\fpbh\sim10^{-3}$ in that mass range. We focus on the standard PBH formation model, where the PBHs are formed through the collapse of the overdensity in the very early Universe. We have review several aspects of PBH formation, and the SIGWs produced from PBHs studied over the past decades. 

First of all, We review the calculation of the PBH formation. The formation of PBHs depends on the power spectrum of the scalar perturbations, the window function, the primordial non-Gaussianities, intrinsic non-Gaussianities and the critical value beyond which PBH can form.

Secondly, we consider the SIGWs inevitably generated during the formation of PBHs. We review the relevant works concerning SIGWs over the past decades. SIGW is a solid prediction of general relativity. But due to the unknown of primordial scalar power spectrum on small scales, the waveform of the SIGW background is model dependent which make the detection of SIGWs very difficult. Fortunately, the logarithm scaling in the infrared region is a model-independent feature for SIGWs, making SIGWs unique from other SGWB. We also review the impacts from the cosmological background, higher-order corrections to SIGWs, primordial non-Gaussianities. Finally, we discuss the gauge issues of SIGWs.
%For a general cosmological background where the equation of state may vary, one has to adopt the numeric method based on \cite{Abe:2020sqb}.
On the other hand, the presence of primordial non-Gaussianities is an essential aspect since it could dramatic change the waveform and the amplitude of the SIGWs. For local-type non-Gaussianities, as long as monochromatic PBH in the allowing mass window represents all the DM, then LISA and PTAs should detect the signals irrespective of $\Fnl$. However, up to $\Gnl$-order, this conclusion no longer holds. The $\Gnl$-order could further suppress the SIGWs than $\Fnl$-order and thus avoiding the detection of LISA. Another important impact of non-Gaussianities is that the power spectrum and the non-Gaussianities are degenerate. As a result, one cannot tell the presence of non-Gaussianities only through the SIGW signal unless considering another independent observation.

So far, SIGW has placed the most severe constraint on $\fpbh$ \cite{Chen:2019xse}. However, this results from monochromatic PBHs in the absence of primordial non-Gaussianities and neglecting the QCD phase transition. During the QCD epoch, the equation of state and the sound speed decrease slightly from $1/3$, thus changing the waveform of SIGWs. This will affect the PBHs within the PTA frequency band. Searching the SIGW signal for realistic models such as considering the QCD phase transition, a specific inflation model, and its loop correction to the power spectrum is necessary in this field.

Over the past decade, SIGWs from PBH DM have been widely studied. SIGW provides a promising tool to verify or falsify the PBH DM hypothesis. We are expected to witness intriguing progress in the coming decades.

	%%%%%%%%%%%%%%%%%%%%%%%%%%%%%%%%%%%%%%%%%%%%%%%%%%%%%%%%%%%%%%%%%%%%%%%%%%%%%%%%
	%%%%%%%%%%%%%%%%%%%%%%%%%%%%%%%% acknowledgments %%%%%%%%%%%%%%%%%%%%%%%%%%%%%%%
	%%%%%%%%%%%%%%%%%%%%%%%%%%%%%%%%%%%%%%%%%%%%%%%%%%%%%%%%%%%%%%%%%%%%%%%%%%%%%%%%
\noindent{\bf{\em Acknowledgments.}}
Cosmological perturbations are derived using the \texttt{xPand} \cite{Pitrou:2013hga} package. We would like to thank Zu-Cheng Chen, Sai Wang, Misao Sasaki and Guillem Dom\`enech for useful discussion. We also acknowledge the use of HPC Cluster of ITP-CAS. 
This work is supported by the National Key Research and Development Program of China Grant No.2020YFC2201502, the grants from NSFC (grant No. 11975019, 11690021, 11991052, 12047503),  the Key Research Program of the Chinese Academy of Sciences (Grant NO. XDPB15), Key Research Program of Frontier Sciences, CAS, Grant NO. ZDBS-LY-7009, and the science research grants from the China Manned Space Project with NO. CMS-CSST-2021-B01.

\noindent{\bf{\em Author contribution.}}
Q.G.H. organizes the whole project. C.Y. performs the main calculations and write the paper.

\noindent{\bf{\em Declaration of interests.}}
The authors declare no competing interests.

\clearpage


\clearpage
	%%%%%%%%%%%%%%%%%%%%%%%%%%%%%%%%%%%%%%%%%%%%%%%%%%%%%%%%%%%%%%%%%%%%%%%%%%%%%%%%
	%%%%%%%%%%%%%%%%%%%%%%%%%%%%%%%%%% references %%%%%%%%%%%%%%%%%%%%%%%%%%%%%%%%%%
	%%%%%%%%%%%%%%%%%%%%%%%%%%%%%%%%%%%%%%%%%%%%%%%%%%%%%%%%%%%%%%%%%%%%%%%%%%%%%%%%
\bibliography{./ref.bib} 
\bibliographystyle{agsm}
	%%%%%%%%%%%%%%%%%%%%%%%%%%%%%%%%%%%%%%%%%%%%%%%%%%%%%%%%%%%%%%%%%%%%%%%%%%%%%%%%
\end{document}